\documentclass[11pt]{article}

\usepackage{amsmath,mathrsfs,amsbsy,amsthm}
\usepackage{cite}
\usepackage{graphicx,epsfig}
\usepackage{latexsym,amssymb}
\usepackage{epsf}
\usepackage{ifpdf,lineno}
\usepackage{color}
\usepackage{geometry}
\def\sech{\hspace{0.75mm}{\rm sech}}
\def\sech{\hspace{0.75mm}{\rm sech}}
	
\begin{document}
	
	\title{{\bf  Kink-Antikink Collisions in the Periodic $\varphi^{4}$ Model}}

\author{M. Mohammadi$^{1}$\thanks{Corresponding Author. } \\
	{\small \texttt{physmohammadi@pgu.ac.ir}}\and
	R. Dehghani$^{2}$ \\{\small \texttt{r.dehghani@queensu.ca}}}
	\date{{\em{$^1$Physics Department, Persian Gulf University, Bushehr 75169, Iran.\\
$^2$ Department of Physics, Engineering Physics $\&$ Astronomy
Stirling Hall, Queen's University 64 Bader Lane Kingston, Ontario, Canada.}}}
	\maketitle

\begin{abstract}

We borrow the   form of potential of the well-known kink-bearing $\varphi^4$ system in the range between its two  vacua and paste  it repeatedly   into the other ranges to introduce   the periodic $\varphi^4$ system. The paper is devoted to providing a comparative numerical study of the properties of the two systems. Although the two systems are quite similar for a kink (antikink) solution, they usually exhibit  different behaviors throughout collisions. For instance, they have different critical velocities,  different results during   collisions, and a different rule in their quasi-fractal structures.   Their  quasi-fractal structures will be studied in the disturbed kink-antikink collisions as well. Hence, three types of  scattering windows will be introduced with respect to the incoming speed, the amplitude, and  initial phase of the internal mode, respectively. Moreover, a detailed comparative study of the collisions  between  two kinks and one antikink will be done at the end.

\textbf{Keywords} : {periodic $\varphi^4$,  kink, soliton, fractal.\\}

\end{abstract}

%	\textbf{Keywords}:{ kink, resonance windows, Periodic $\varphi^4$ Model.}
	
%	\section*{Contents}
%	\subsubsection*{1. Introduction............................1}
 %   \subsubsection*{2. The Periodic $\varphi^{4}$ Model............2}
  %  \subsubsection*{3. The Resonance Windows.........3}
   % \subsubsection*{4. Numerical Results....................4}
    %\subsubsection*{5. Conclusion................................6}
    %\subsubsection*{Refrences......................................7}

	\section{Introduction}\label{sec1}

Nonlinear field models  with topological kink (antikink) solutions in $1 + 1$ dimensions
are of growing interest for theoretical physics from high energy physics and cosmology to
condensed matter physics \cite{CO1,CO2,CO3,CO4,CO5,CO6,CO7,CO8,DW1,DW2,DW3,DW4,DW5}.
Especially in cosmology, the structure and dynamics of  domain walls, can be modeled or  described  by the ($1+1$)-dimensional kink-bearing   theories  \cite{DW1,DW2,DW3,DW4,DW5,DW6,DW7}.  Topological kink (-like) solutions also exist in more complex models with two or more fields in ($1 + 1$)-dimensions \cite{CU1,CU2,CU3,CU4,CU5,CU6,CU7,CU8,CU9,CU10,CU11,CU12,CU13}. Complex kink (antikink) solution is another type of topological soliton-like solutions which was obtained  for a complex nonlinear  Klein-Gordon field system \cite{CK}.

%time-consuming computation

	The dynamic  and other properties  of kinks have been  of great
importance and have attracted the attention of physicists and mathematicians for a long
time [30-89]. In particular, the kink-(anti)kink scattering and the interactions of kinks with impurities were actively studied previously \cite{IM1,IM2,IM3,IM4,IM5,IM6,IM7,IM8,IMPUr}. In this context, the recent interesting  results on kink-antikink interactions in models, which possess kinks with power-law tails, can be mentioned \cite{PLT1,PLT2,PLT3,PLT4,PLT5,PLT6}. It is also worth mentioning that the recent  results on  the study of maximal values of different quantities in multi-soliton (kink and antikink) collisions have been another topic of interest to researchers in recent years \cite{EV1,EV2,EV3}. There have been different methods to study the behaviours  of   kinks (antikinks) in the interactions among which one can mention the quasi-exact numerical methods, and the approximate methods such as the collective coordinate approximation \cite{EZ10,CC0,CC1,CC2,CC3,CC4,Man3} and the Manton's method \cite{Man1,Man2,Man3}. However, in this paper, we only use a numerical method to obtain the results.

%the $\phi^8$ and more complex models \cite{PHI81,PHI82,PHI83,PHI84}.
%Moreover,

For some kink solutions, there is an  ability  to keep  a constant   oscillatory internal motion with a specific frequency. This phenomenon is related to whether there is a non-trivial internal mode for the kink (antikink) solution \cite{INTER1,FR2,INTER2}. Internal modes are the bound states of a Schr\"{o}dinger-like equation  which was obtained by considering the  small  fluctuations on a kink solution. Kink-bearing systems, depending on whether they have  non-trivial internal modes, can be divided into nearly integrable and non-integrable models \cite{INTER2,AI1,AI2}. In this context the only integrable model is the well-known sine-Gordon (SG) system. It was shown that the energy loss due to the radiation during the collision is usually small in the nearly integrable models in comparison with non-integrable models. The amount of radiation is a complicated function of the initial speed, and depending on that, the fate  of a kink-antikink collision can  be completely  different.
In general, the collision between a kink and an antikink may lead to a long-living non-topological oscillating bound state, so called a bion  state, or they  may eventually bounce back and reflect from each other, or they may  annihilate immediately  in  radiative systems \cite{RD}.

For any  kink-bearing system except the SG system, there is always a critical speed $v_{cr}$ for which if  the initial speed $v_{in}$ of a head-on  kink-antikink collision  is greater, kink and antikink pass through one another and reappear after collisions with a constantly vibrational behavior.
If the initial  incoming speed $v_{in}$ is smaller than  the critical speed $v_{cr}$ $(v_{in}< v_{cr})$, the kink and antikink usually form  a bion state  that decays slowly and radiates energy in the form of small-amplitude waves.
In a number of models, when $v_{in}< v_{cr}$, there have been spotted   a new interesting phenomenon called the  escape or scattering  windows. For such initial speeds, because of the resonance energy exchange between the translational and vibrational internal  modes, the two kinks (antikinks) will not form a bion and will bounce off each other after two or more collisions \cite{EZ00,EZ4,TV1,TV2}.
Moreover, a prominent feature of such systems is the appearance of a  chaotic quasi-fractal structure  with a hierarchical  order of scattering windows \cite{FR2,FR1,FR3,FR4}.

\begin{figure}[h!]
	\centering
	\includegraphics[scale=1]{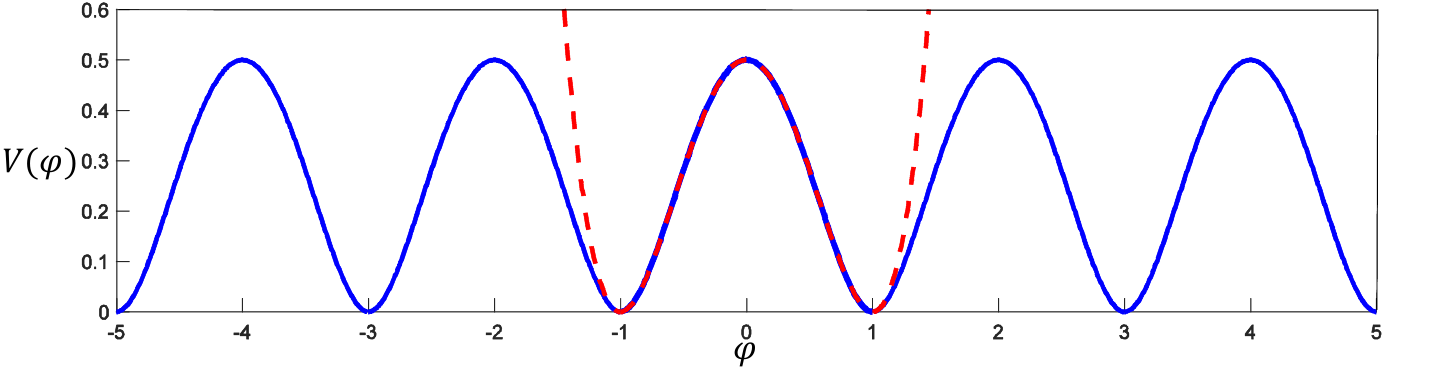}
	\caption{ The red dash (solid blue) curve is  representing the potential of the (periodic) $\varphi^4$ model. In fact  we copy the potential of the  $\varphi^4$ model in the range  from $-1$ to $1$ and paste it in multiple regions such as $-1$ to $-3$, $-3$ to $-5$, $1$ to $3$, $3$ to $5$, and so on, to introduce the  potential  of the periodic $\varphi^4$ system.  }\label{Pot}
\end{figure}

In this regard, the  $\varphi^4$ model which   is a well-known kink-bearing system,  was studied extensively, namely, in relation to the resonant  kink-antikink scattering and the quasi-fractal structure \cite{EZ00,FR2,FR1}, kinks interaction
with impurities  \cite{IM4,IM5,IM6,IMPUr}, high energy density in the collision of $N$ solitons  \cite{EV2},  ac external force \cite{PHI41}, scattering between wobbling kinks \cite{newg},  and  the periodically modulated
on-site potential \cite{PHI42}.
 The  corresponding  potential of the $\varphi^4$ system is as follows:
\begin{equation}\label{ffg}
V(\varphi)=\frac{1}{2}(\varphi^2-1)^2,
\end{equation}
 It has a single non-trivial internal mode $\psi(x)\varpropto \tanh(x)\sech(x)$  with a specific  rest frequency $\omega_{o}=\sqrt{3}$  \cite{EZ00,FR2}.  In this paper, inspired by the well-known $\varphi^4$ model,  a new kink-bearing system can be introduced  that can be called the periodic $\varphi^4$ model with following form of the potential \cite{Azizi}:
 \begin{equation}\label{eq4}
  V(\varphi)=\frac{1}{2}((\varphi-2N)^2-1)^2, \quad 2N-1\leqslant\varphi<2N+1,
 \end{equation}
where $N=0,\pm1,\pm2,\cdots$.
The  potential of the new system is the same as that of the  $\varphi^4$ system in the range $-1<\varphi<1$, which is repeated  in  other regions of the real scalar field $\varphi$ (see Fig.~\ref{Pot}).
 Although both systems have  the same form of potential in the range $-1\leq \varphi\leq 1$ and have identical  soliton (kink and antikink) solutions,  they exhibit different behaviors  in the collisions (interactions) due to their differences elsewhere  (i.e. $\varphi<-1$ and $1<\varphi$).  From a physical point of view, this is important because it attracts attention
 to fact that  the similarity of particles may not necessarily mean that their interaction is the same. In other words, comparing these two particular models, just as an example, indicates the possibility that there may be similar particles in the nature that exhibit different interaction behaviors.    The main purpose of this paper is to present a comparative study of the interaction properties of these systems.
 Accordingly,  the collisions,  the scattering  windows, and the quasi-fractal structure of both systems  will be investigated and compared in detail.

%In other words,
%but for the other ranges with the same field spacing $2$ from $1$  ($-1$) to infinity (-infinity), the same basic form of the potential in the range  $-1<\varphi<1$  repeat for all of them, i.e.

%These differences are originated  from the different  potentials in the ranges $\varphi<-1$ and $\varphi>1$, accordingly we can call  the  potential in this ranges '\textit{the interaction potential}''.

%In particular, it was  shown numerically that the critical speed $v_{cr}$ for this new system is smaller than that of the ordinary $\varphi^4$ system. The collisions of  the kink-antikink  pairs, in this  new model, lead to lower amplitudes of the   oscillations on which they can pass through one another. Moreover, the amount of the radiative energies in any kink-antikink collision are smaller in comparison with that of the ordinary $\varphi^4$ system. To sum up,  the behavior of the periodic $\varphi^4$ system during collisions is closer to an integral system.

Our paper is organized as follows. In section  \ref{sec2}, we review,
briefly, the formulation and some general  properties of a relativistic kink-bearing system in $1+1$ dimensions. Also, the necessary numerical considerations for  the obtained  results  are presented in this section.
 In Section
\ref{sec3}, the internal modes of the kink (antikink) solutions are considered in detail.  Section \ref{sec4} is devoted to all numerical results, which have been  obtained for  both $\varphi^4$ and periodic $\varphi^4$ systems. The last section is devoted to  summary and conclusions, where we also formulate some possible directions for further research.

\section{Basic Equations}\label{sec2}

%In this paper we study the behaviours of kink and antikink solutions of  a new potential that can be called the periodic $\varphi^4$ system (\ref{eq4}).
% In fact, we borrow the  potential form  of the well-known $\varphi^{4}$ system in the interval   $-1\leq\varphi<1$, and use  it for other intervals  $ 2n-1\leqslant\varphi<2n+1$ (see Fig.~\ref{Pot}).

In general, the Lagrangian density  of a kink-bearing system in $1+1$ dimensions   is
 \begin{equation}\label{21}
 {\cal L}=\frac{1}{2}\left(\frac{\partial\varphi}{\partial t}\right)^{2}-\frac{1}{2}\left(\frac{\partial\varphi}{\partial x}\right)^2-V(\varphi),
 \end{equation}
where $\varphi$ is a real scalar field and the self-interaction term $V(\varphi)$ is called the potential.
Using the Euler-Lagrange equation, the equation of motion can be derived as:
 \begin{equation}\label{eq3}
 \ddot{\varphi}-\varphi''=-\frac{dV}{d\varphi},
 \end{equation}
where $\ddot{\varphi}$ and ${\varphi''}$ are the second derivatives of the scalar field ${\varphi}$ with respect to time and space, respectively.
For the dynamical field equation (\ref{eq3}), there are various manifestations of the potential $ V(\varphi)$ that yield well-known kink (antikink) solutions. In fact, if the positive definite  potential $ V(\varphi)$ has at least two degenerate vacua (i.e. points of minimum
potential), there will be localized solutions called kinks and antikinks with positive
and negative topological charges, respectively. For the $\varphi^4$ model (\ref{ffg}), there are only two degenerate minima  (vacua),  at $-1$ and $1$, thus, there is only one type of kink and antikink solution that  belongs to a unique sector $(-1,1)$ \cite{EZ0}. However, for the periodic $\varphi^4$ model (\ref{eq4}), similar to the well-known sG model,  there are  infinite  vacua, i.e. any odd number (see Fig.~\ref{Pot}), hence there are infinite types of kink and antikink solutions belonging  to infinite  sectors $(2N-1,2N+1)$.

In order to find a moving non-vibrational   topological kink solution,  we should consider  the  dynamical equation (\ref{eq3}) for a solution in the following form:   $\varphi_{v}=\varphi_{o}(\gamma(x-x_{o}-vt))$, where $v$ is the velocity of the kink, $\gamma=1/\sqrt{1-v^2}$ is the Lorentz factor, $x_{o}$ is the initial position, and $\varphi_{o}$  is an unknown function which should  be found.  If one does this procedure   for the periodic $\varphi^4$ system (\ref{eq4}), the moving non-vibrational kink and antikink solutions for sector $(2N-1,2N+1)$, i.e. $2N-1< \varphi< 2N+1$, will be
\begin{equation}\label{5}
\varphi_{v}(x,t)=\tanh(\pm\gamma(x-x_{o}-vt))+2N,
\end{equation}
where $+$ ($-$) is used for kinks (antikinks),   and $N$ is any integer number.  Note that the above solutions for $N=0$ are the same  kink and antikink solutions of the ordinary $\varphi^4$ system.

For such a system,  the superposition of  two or multiple kinks and antikinks  can be assumed as new solutions of the system, provided they are  far enough from one another. For example, for  $m$  number  of the solitary wave solutions (kinks and  antikinks), which initially have different  velocities $v_{i}$ and  initial positions $x_{i}$, the following combination
\begin{equation}\label{7}
\varphi=\sum_{i=1}^{m}\tanh(\pm\gamma_{i}(x-v_{i}t-x_{i}))+C, \quad x_{i+1}-x_{i}\gg1,
\end{equation}
where $\gamma_{i}=1/\sqrt{1-v_{i}^2}$,  is again a  solution of the system. Here, $\pm$ means that for any solitary  wave solution which  initially stands at $x_{i}$, choosing $+$ (kink) or $-$ (antikink) is optional. The constant  $C$ is a proper number  which should be included in order to have right boundary conditions.
In fact, the relative distance between the kinks and antikinks are quite large to ensure that the overlap of the kinks and antikinks are negligibly small.
It should be noted  that, for the ordinary $\varphi^4$  system, since there are  two vacuum points at $-1$ and $+1$,  only  the collisions  of   the alternative combinations of the kink and antikink  solutions are possible to be studied. However, for the periodic  $\varphi^4$ model  there are no  conditions on the  initial  arrangement of kinks and antikinks  like the SG system.

In general, since it has not been possible to obtain multisolitonic solutions of the non-integrable  systems analytically,  it is common to use numerical methods to study the  collisions of  any number of  kinks and antikinks. Using a superposition of several far apart  kinks and antikinks, which are moving towards the collision point,   is the necessary initial condition  for a  numerical investigation of the collisions.     To acquire numerical results of the equation of motion (\ref{eq3}), we use the discretized version of that in the following form \cite{FR1}:
\begin{equation}\label{dd}
\frac{\partial^2 \varphi_{n}}{\partial t^2}-\frac{1}{h^2}(\varphi_{n-1}-2\varphi_{n}+\varphi_{n+1})+\frac{1}{12h^2}(\varphi_{n-2}-4\varphi_{n-1}+6\varphi_{n}-4\varphi_{n+1} +\varphi_{n+2})+\dfrac{dV(\varphi_{n})}{d\varphi_{n}}=0,
\end{equation}
where $V$ is represented either by Eq.~(\ref{ffg}) or Eq.~(\ref{eq4}),  $h$ is the small  spatial step, $n=0,\pm 1,\pm 2,\cdots$ and $\varphi_{n}=\varphi(nh,t)$.
A fourth-order Runge-Kutta scheme with the small time-step   $k$ is used to solve the ordinary differential equations (\ref{dd}) numerically. The accuracy of this  standard method is to fourth order in both temporal  and spatial steps. In this paper, all the simulations were
carried out for  $h=k=0.02$. To avoid the reflective effects of  boundaries on the accuracy of simulations, we fix them at far  distances from the origin ($x=0$), namely  from $-200$ to $200$ in this paper. It is also necessary to say that all simulations were done in the time interval $0<t<400$.

From the Noether's theorem, the energy functional corresponding to the Lagrangian (\ref{21}) is viewed as:
\begin{equation}\label{9}
E[\varphi]=\int_{-\infty}^{+\infty}\varepsilon(x,t) ~dx=\int_{-\infty}^{+\infty}\left(\frac{1}{2}\dot{\varphi}^{2}+\frac{1}{2}\varphi'^{2} +V(\varphi)\right)dx=K+U+P.
\end{equation}
where
\begin{equation}\label{11}
  \varepsilon(x,t)=k(x,t)+u(x,t)+p(x,t),
\end{equation}
is the energy density function  and functions $k(x,t)$, $u(x,t)$, and $p(x,t)$ are introduced as
  \begin{equation}\label{12}
   k(x,t)=\frac{1}{2}\left(\frac{\partial\varphi}{\partial t}\right)^2,\quad
     u(x,t)=\frac{1}{2}\left(\frac{\partial\varphi}{\partial x}\right)^2,\quad
     p(x,t)=V(\varphi)=\frac{1}{2}((\varphi-2n)^{2}-1)^{2}
\end{equation}
Accordingly, the total energy of the system (\ref{9}) can be written as the sum of three portions: the kinetic energy $K$, the gradient energy $U$, and the potential energy $P$, that are defined as the integrations of the $k(x,t)$, $u(x,t) $ , and $p(x,t) $, above the whole space, respectively. Hence, $k(x,t)$, $u(x,t) $ , and $p(x,t)  $ are called the kinetic,  the gradient, and the potential energy density, respectively. The details of any collision can be more clarified by studying the evolution of all these parts throughout the collisions.

In the numerical calculations, we need to somehow be able to obtain the velocity of an entity after the collisions.
 To do that, we calculate the energy $E$ (\ref{9}) and the momentum $P$ of this entity and simply use the relativistic relation $v=P/E$.
 The corresponding momentum for the Lagrangian density (\ref{21}) would be:
\begin{equation}\label{bng}
P[\varphi]=\int_{-\infty}^{+\infty}(-\dot{\varphi}\varphi') dx,
\end{equation}
that is another  obvious  result from the Noether's theorem as well as  equation (\ref{9}).

\section{Internal modes}\label{sec3}

%In other words, the  energy of a  deformed kink (antikink) solution is always larger than the rest energy  of the non-deformed one.

A kink (antikink)  has the lowest energy among the other solutions with  the same  asymptotic behaviour \cite{EZ0,Derrick}. Therefore,  we can expect that any   permissible small deformation above   a  kink (antikink) solution, finally leads to an increase in the total energy.  In general, a small   deformed kink solution which is at rest,   can be introduced as follows:
\begin{equation} \label{defo}
\phi_{o}(x,t)=\varphi_{o}(x)+\delta \varphi(x,t),
\end{equation}
 where $\varphi_{o}(x)$ is the same non-moving kink solution (for example Eq.~(\ref{5}) for $v=0$) and $\delta \varphi(x,t)$  is  any permissible small function. Note that a permissible deformation $\delta \varphi(x,t)$  is one for which $\phi_{o}(x,t)$ is again a  solution  of the equations of motion (\ref{eq3}). In other words, for a non-moving  kink (antikink) solution which is slightly deformed (\ref{defo}),  we expect:
 \begin{equation} \label{eqd}
\Box (\varphi_{o}+\delta \varphi)=(\delta \ddot{\varphi})-(\varphi_{o}''+\delta \varphi'')=-\dfrac{dV(\varphi_{o}+\delta \varphi)}{d(\varphi_{o}+\delta \varphi)},
 \end{equation}
 note that  $\ddot{\varphi_{o}}=0$. From Eq.~(\ref{eq3}), for a non-moving kink solution we have: $\varphi_{o}''=\frac{dV(\varphi_{o})}{d\varphi_{o}}$. Therefore,  expanding to
the first order in $\delta\varphi$, Eq.~(\ref{eqd}) simplifies  to
 \begin{equation} \label{seqd}
\Box (\delta \varphi)=\delta \ddot{\varphi}-\delta \varphi''\approx -{\cal U}(x)\delta\varphi,
 \end{equation}
where
${\cal U}(x)=\frac{d^2U(\varphi_{o})}{d\varphi_{o}^2}$
can be called the “\emph{kink potential}'', it is also called “\emph{stability potential}" or
“\emph{quantum-mechanical potential}".
 Equation (\ref{seqd}) can be considered as the dominant dynamical equation for the small permissible deformations  $\delta \varphi$. Since Eq.~(\ref{seqd}) is a   linear homogenous partial differential equation, we can solve it using variables separation method. Hence, one can consider a solution of the form $\delta\varphi(x,t)=\psi(x)\chi(t)$, provided $\psi(\pm\infty)=0$, and substitute back into Eq.~(\ref{seqd}). Finally, it leads to two independent ordinary differential equations:
 \begin{eqnarray}\label{sdd}
&& \ddot{\chi}=-\omega_{o}^2\chi,
\\&&\label{sdd2}
 -\psi''+{\cal U}(x)\psi=\omega_{o}^2\psi,
 \end{eqnarray}
 where $\omega_{o}^2$ is the constant of separation.  The trivial independent solutions of Eq.~(\ref{sdd}) are $\cos(\omega_{o}t)$ and $\sin(\omega_{o}t)$, respectively. Equation (\ref{sdd2}) is a Schr\"{o}dinger-like  equation for which there are two  different  types  of solutions which are called   internal modes (bound states) and free modes. Internal modes (free modes) are some discrete (continuous) solutions of Eq.~(\ref{sdd2}) for which $\omega_{o}^2<{\cal U}(\pm \infty)={\cal U}_{f}$ ($\omega_{o}^2>{\cal U}(\pm \infty)={\cal U}_{f}$) and exponentially  (periodically) tend to zero  ($\sin(\sqrt{\omega_{o}^2-{\cal U}_{f}}~x)$ or $\cos(\sqrt{\omega_{o}^2-{\cal U}_{f}}~x)$) at large distances.

In general,  Eq.~(\ref{sdd2}) always has a trivial solution  $\psi=\xi\dfrac{d\varphi_{o}}{dx}$ with $\omega_{o}=0$, where $\xi$ is  arbitrary provided that $|\psi|\ll 1$.  However,  this trivial solution  is
associated only with an infinitesimal translation of the static kink (antikink)-solution \cite{EZ1,FR2,EZ2}, it has no other physical meaning.
It has been  seen  numerically  that the kink   solutions,  which have non-trivial  bound states (internal modes), can   keep a
constantly oscillating behaviour after collisions (something similar to what is  seen in  Figs.~\ref{p3} and \ref{pp3}). For example, for the  $\varphi^4$ (periodic  $\varphi^4$) system, which was introduced in the previous section, the related kink potential is ${\cal U}(x)=4-6\sech^2(x)$ for which there is a non-trivial bound state (internal mode) corresponding to $\omega_{o}^2=3$ and $\psi\propto \tanh(x)\sech(x)$ \cite{EZ1,FR2}.
The other systems, which have  no non-trivial bound states, can never maintain a constantly  oscillating behaviour  after the collisions \cite{AI2,CU8,EZ2}. In fact, any non-trivial  internal mode can be considered as a channel to impose an additional  fluctuation   on the kink (antikink) solution. To put it differently,  it is a channel for kink (antikink) solution to absorb some external energies.

Relativistically speaking, if we know the exact space-time function of a scalar field in the rest frame, using a boost, we can get the corresponding space-time function in other inertial frames as follows: $t\longrightarrow \gamma(t-vx)$ and $x\longrightarrow \gamma (x-vt)$. For example, based on  pervious discussions, if one considers a kink solution of  both  $\varphi^4$ and periodic $\varphi^4$  systems,   the general
disturbed version of that at rest can be introduced in the following form:
\begin{equation} \label{fff}
\phi_{o}(x,t)=\varphi_{o}(x)+\psi(x)\sin(\omega_{o}t+\theta_{o}),
\end{equation}
where $\omega_{o}^2=3$, $\psi(x)=\xi\tanh(x)\sech(x)$ is  the  single non-trivial  bound state of Eq.~(\ref{sdd2}), and $\theta_{o}$ is  an arbitrary initial phase.  Therefore, the moving version of this disturbed kink (antikink) solution can be obtained easily:
\begin{equation} \label{ggg}
\phi_{v}(x,t)=\phi_{o}(\gamma(x-vt),\gamma(t-vx))=\varphi_{o}(\gamma(x-vt))+\psi(\gamma(x-vt))\sin(\omega_{o}\gamma(t-vx)+\theta_{o}).
\end{equation}
by introducing $\omega=\omega_{o}\gamma$ and $k=\omega_{o}\gamma v$, we can simplify Eq.~(\ref{ggg}) to
\begin{equation} \label{ggf}
\phi_{v}(x,t)=\varphi_{v}(x,t)+\psi(\gamma(x-vt))\sin(\omega t-kx+\theta_{o}),
\end{equation}
where $\varphi_{v}(x,t)=\varphi_{o}(\gamma(x-vt))$ is the same  moving undisturbed kink (antikink) solution.

\section{Collisions}\label{sec4}

In this section, we mainly focus on the results of the collisions between a kink and an antikink of both systems and compare them with one another. We set the conditions such as initial positions and velocities so that the kink and the antikink simultaneously arrive at a special  point in space (i.e. $x=0$)  and collide.
The relative distance between the kink and antikink is quite large to ensure that the overlap of the kink and antikink is negligibly small. For a comparative study between the two systems, the initial and boundary    conditions must be the same for both. Hence, since two systems have the same potential in the range  $-1\leq \varphi \leq 1$, the combination  of the initial  kinks and anti-kinks, that their collision we are going to study, should be in such a way that the initial field $\varphi$ is  between $-1$ and $1$.

\subsection{Primary Results}

In a kink-antikink collision, due to the symmetrical potential of the  $\varphi^4$ (periodic  $\varphi^4$) model, there are no differences between the results of  different orientations such as kink-antikink ($K\overline{K}$) and antikink-kink ($\overline{K}K$).
The initial condition for this collision  in the  $\varphi^4$ (periodic  $\varphi^4$) model is:
\begin{equation}\label{fkk}
\varphi_{K\overline{K}}=\tanh(+\gamma(x-vt-a))+\tanh(-\gamma(x+vt-b))-1,
\end{equation}
where $a=-20$ and $b=20$ are the initial positions, and $v$ is the  incoming speed.  In the case of a kink-antikink collision, there is a critical speed which  separates two regions of incoming speeds. For  speeds less  than the critical speed,  kink and antikink  usually   stick together and generate a bion state. For the $\varphi^4$ system, the critical speed is  about $v_{cr}=0.2600$, and for the periodic $\varphi^4$ we have found $v_{cr}$ to be about $0.1516$.  This difference  arises  from the potential differences in the ranges $\varphi<-1$ and $\varphi>1$. In other words, although two systems have the same kink and antikink solutions, their differences in the interaction region, i.e. $\varphi<-1$ and $\varphi>1$,  cause different critical speeds.

\begin{figure}[ht!]
   \centering
   \includegraphics[width=150mm]{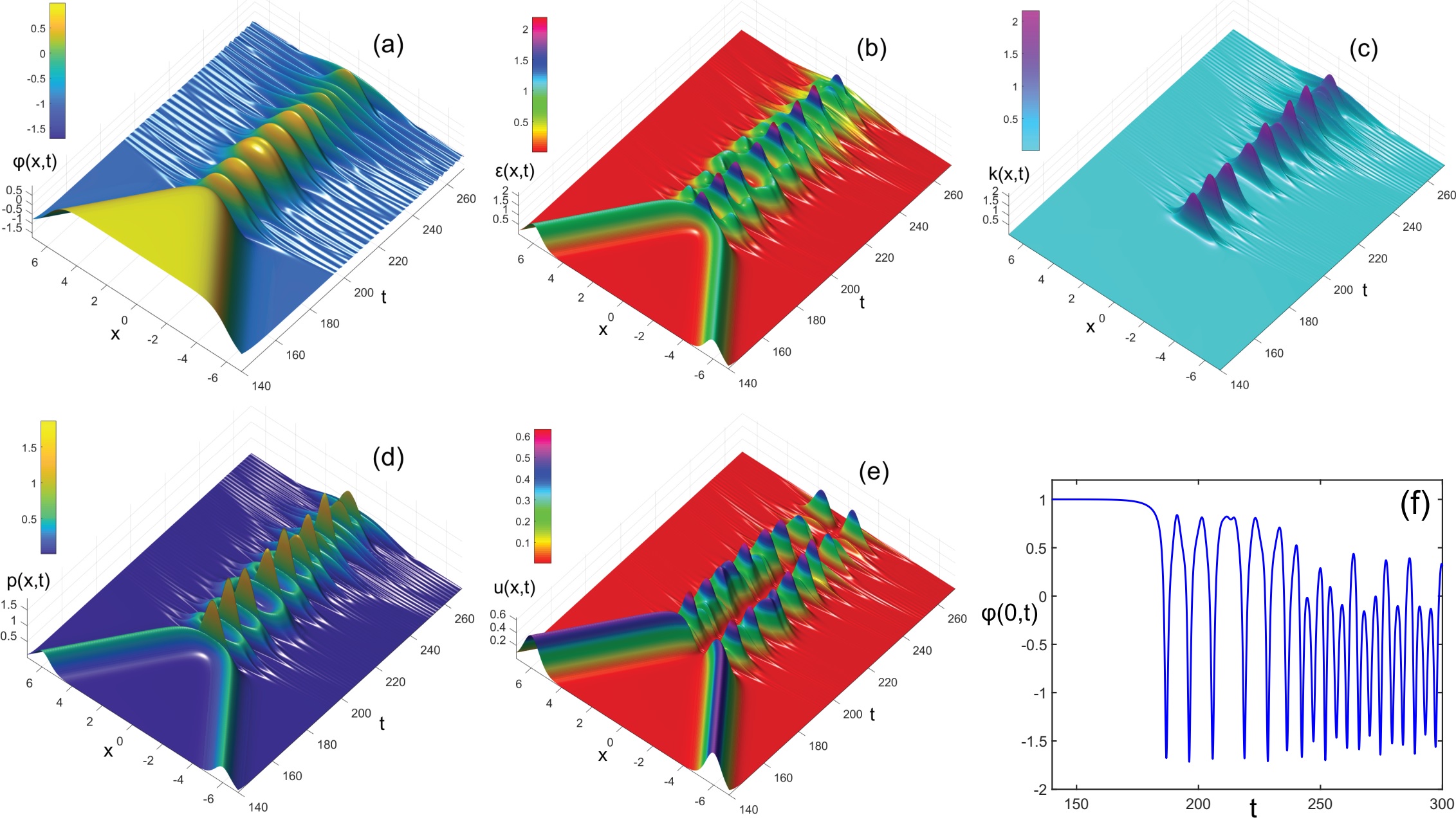}
   \caption{Some details of a kink-antikink  collision in the context of  $\varphi^4$ system with initial speed $v=0.1$. Figures $a$-$f$  represent  the variation of the field, energy density, kinetic energy density, potential energy density, gradient energy density, and the field at center of mass, respectively. Note that, these explanations are similar in the following three figures and will not be repeated.  } \label{p1}
 \end{figure}

\begin{figure}[ht!]
   \centering
   \includegraphics[width=150mm]{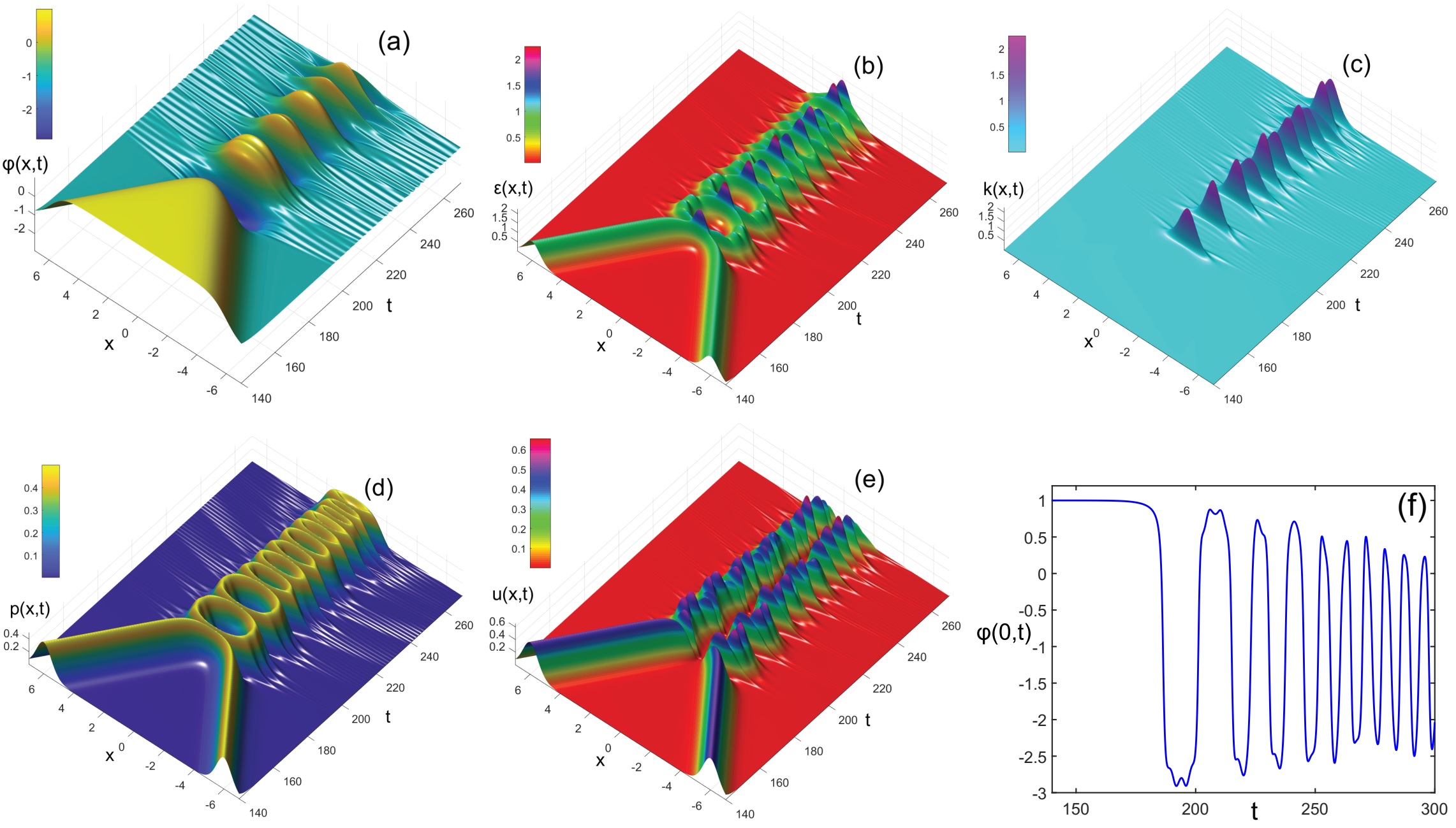}
   \caption{Some details of a kink-antikink  collision in the context of periodic  $\varphi^4$ system with initial speed $v=0.1$.  } \label{pp1}
 \end{figure}

A bion is a long-living bound state with zero topological charge, which decays slowly via emitting its energy in the form of small amplitude waves.        Figure \ref{p1} (\ref{pp1}) shows  some  details of a $K\overline{K}$ collision at the same initial speed $v=0.1$  below  its critical speed in the context of the  $\varphi^4$ (periodic  $\varphi^4$) system, which finally leads to the generation of a bion state.
From the numerical analysis we can extract the extreme  values for the $\varphi^4$ system:
  \begin{equation}\label{dcvb}
 k_{max}=2.1642, \quad u_{max}= 0.6332, \quad p_{max}=1.8832, \quad \varepsilon_{max}=2.1902.
 \end{equation}
Moreover, for the periodic $\varphi^4$ system we obtain:
\begin{equation}\label{dcvb2}
 k_{max}=2.2452, \quad u_{max}=0.6707, \quad p_{max}=0.5000, \quad \varepsilon_{max}=2.2556.
 \end{equation}

\begin{figure}[ht!]
   \centering
   \includegraphics[width=150mm]{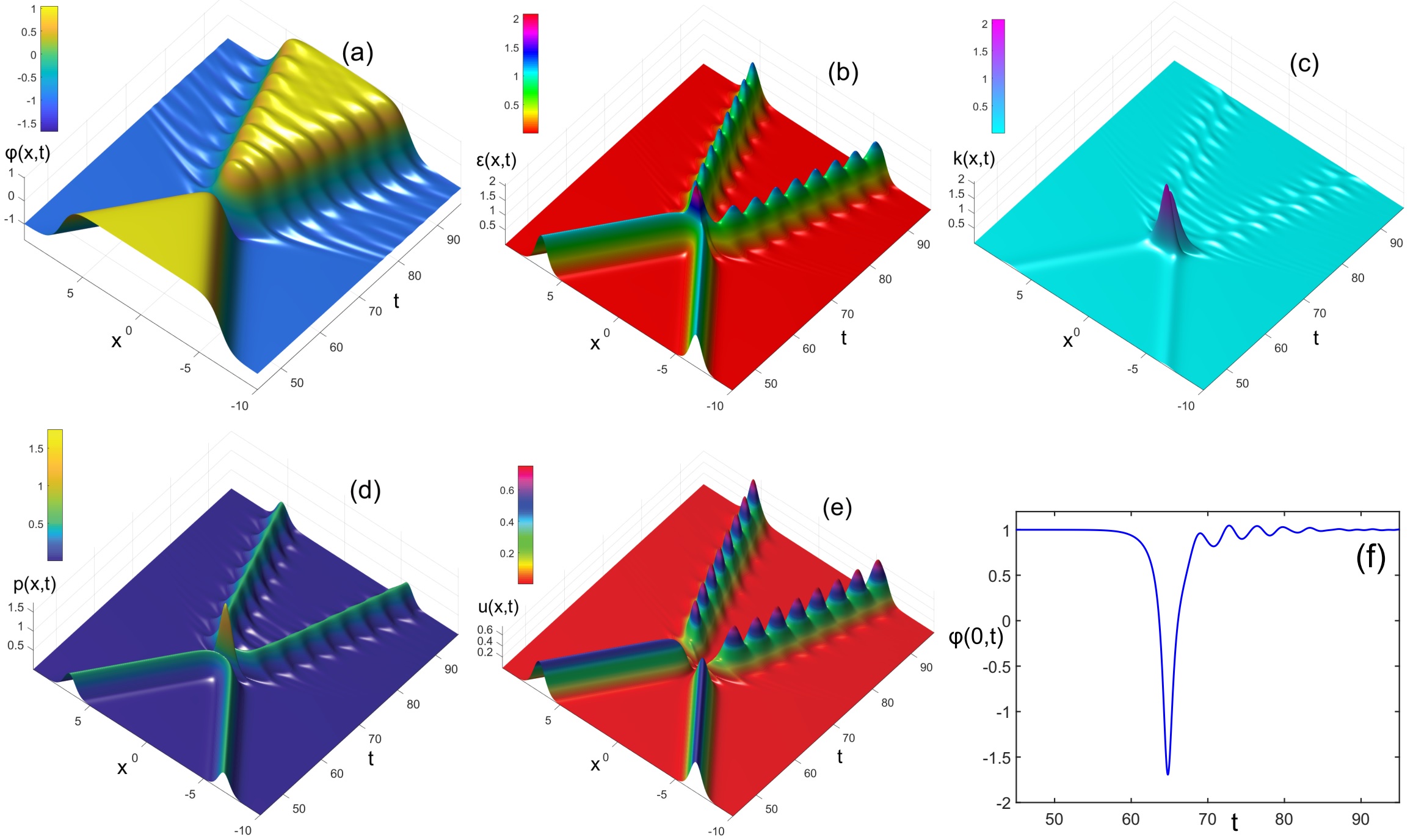}
   \caption{The details of a kink-antikink  collision in the context of  $\varphi^4$ system with initial speed $v=0.3$. } \label{p3}
 \end{figure}

\begin{figure}[ht!]
   \centering
   \includegraphics[width=150mm]{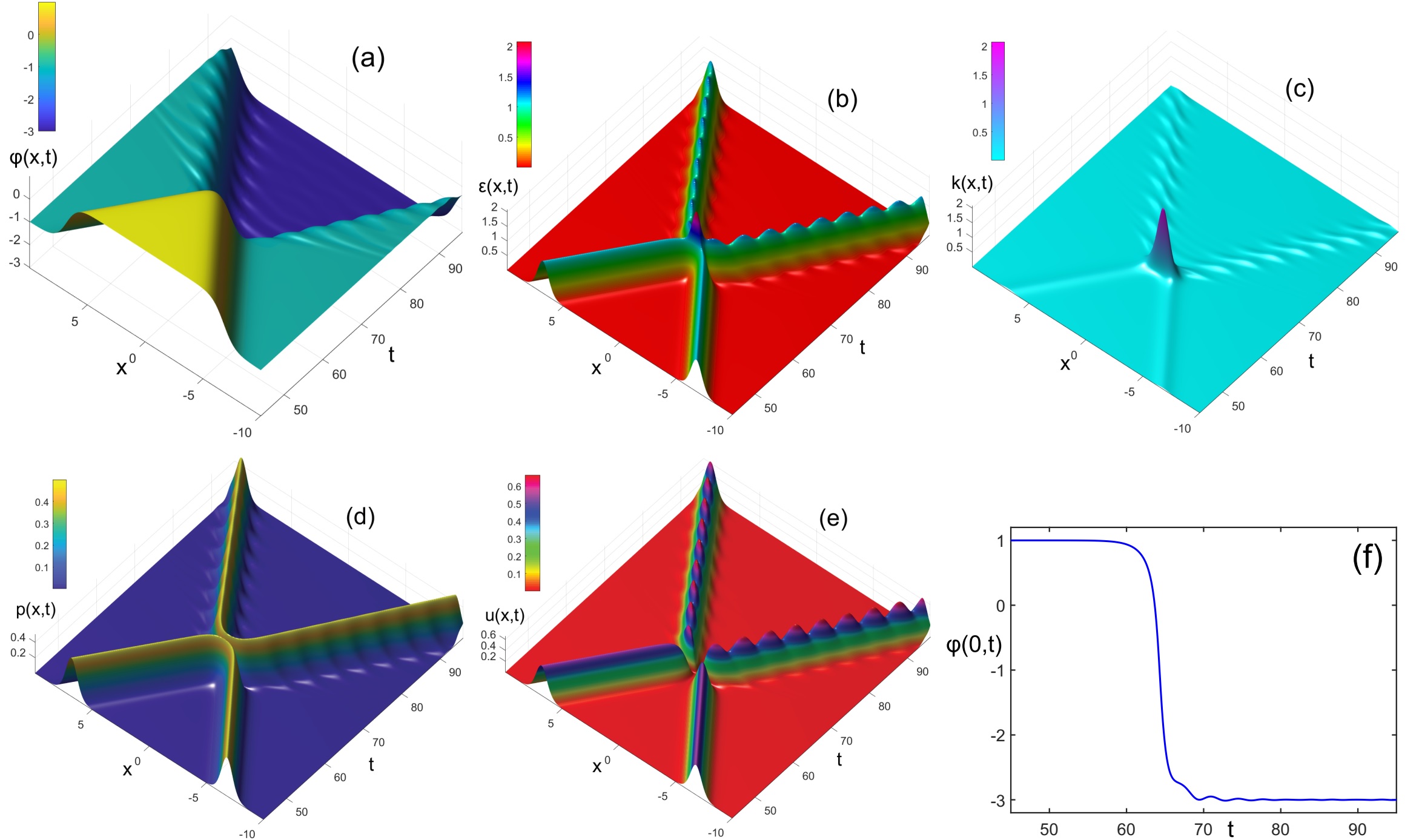}
   \caption{The details of a kink-antikink  collision in the context of periodic $\varphi^4$ system with initial speed $v=0.3$.  } \label{pp3}
 \end{figure}

For  initial speeds larger than the critical speed (i.e. $v>v_{cr}$), kink and antikink always escape from each other after  collisions. As an example, some details  for such  collisions with  $v=0.3$ are  shown in Figs.~\ref{p3} and \ref{pp3} for the $\varphi^4$ and periodic $\varphi^4$  systems, respectively.
 Furthermore, the extreme values for the  $\varphi^4$ system  are:
\begin{equation}\label{bcvb}
k_{max}=2.0751, \quad u_{max}=0.7530, \quad p_{max}=1.7482, \quad \varepsilon_{max}=2.0844.
\end{equation}
and for period $\varphi^4$ system:
\begin{equation}\label{bcvb2}
 k_{max}=2.0753, \quad u_{max}=0.6623, \quad p_{max}=0.5000, \quad \varepsilon_{max}=2.0844.
 \end{equation}

Examining the previous four figures, one realizes that  except the potential energy density $p$,  the evolution of $k$, $u$ and $\varepsilon$ are almost similar for the two systems.
Moreover, comparing  Eq.~(\ref{dcvb}) with  Eq.~(\ref{dcvb2}) and also comparing  Eq.~(\ref{bcvb}) with  Eq.~(\ref{bcvb2}) shows that  the values of $p_{max}$ differ substantially. According to Eq.~(\ref{12}),  $p$ is the same potential $V(\varphi)$, and the potential of the periodic $\varphi^4$ system is confined to values less than $V(2n)=0.5$ ($n=0,\pm1,\pm2,\pm3,\cdots$), whereas the potential of the $\varphi^4$ system is not confined. In fact, for  $\varphi>\sqrt{2}$, the potential of the  $\varphi^4$ system is larger than $0.5$. In a collision process, according to part $f$ in Figs.~\ref{p1}-\ref{p3}, the field $\varphi$ changes and increases (decreases) to amounts larger  (smaller) than $\sqrt{2}$ ($-\sqrt{2}$). Hence,   the maximum  value  of $p$ in the  $\varphi^4$ system would be larger than $0.5$, but for the periodic $\varphi^4$ system,  it would be   $0.5$.   Furthermore, numerical analysis shows that for  initial speeds larger than the critical speed of the $\varphi^4$ system, i.e. $v>0.2600$, the extreme values of the energy density function are approximately the same in both systems. However, unlike the $\varphi^4$ system, for $v>v_{cr}$, a $K\overline{K}$ collision in the  periodic $\varphi^4$ system, always leads to a pair of the $\overline{K}K$. Meanwhile, as Figs.~\ref{p3} and \ref{pp3} demonstrate,   the output  speed as a function of the incoming speed in the  $\varphi^4$ system is always smaller than that of the periodic $\varphi^4$ system. Moreover, the amplitude of the induced vibrations after collisions, which are originated from  internal modes,  are smaller in the periodic $\varphi^4$ system.

 \subsection{The Resonance Windows}

In general, for a  kink-antikink collision, usually one of the three following situations will occur: Either they stick together and generate a bion state, or that they do not even feel each other's influence and get past each other having initial velocities near the speed of light.  The third situation is that they bounce back and reflect from each other. For some systems, so-called radiative systems \cite{RD}, there is another special situation in which the kink-antikink pair will be annihilated immediately after  collision.

\begin{figure}[ht!]
   \centering
   \includegraphics[width=140mm]{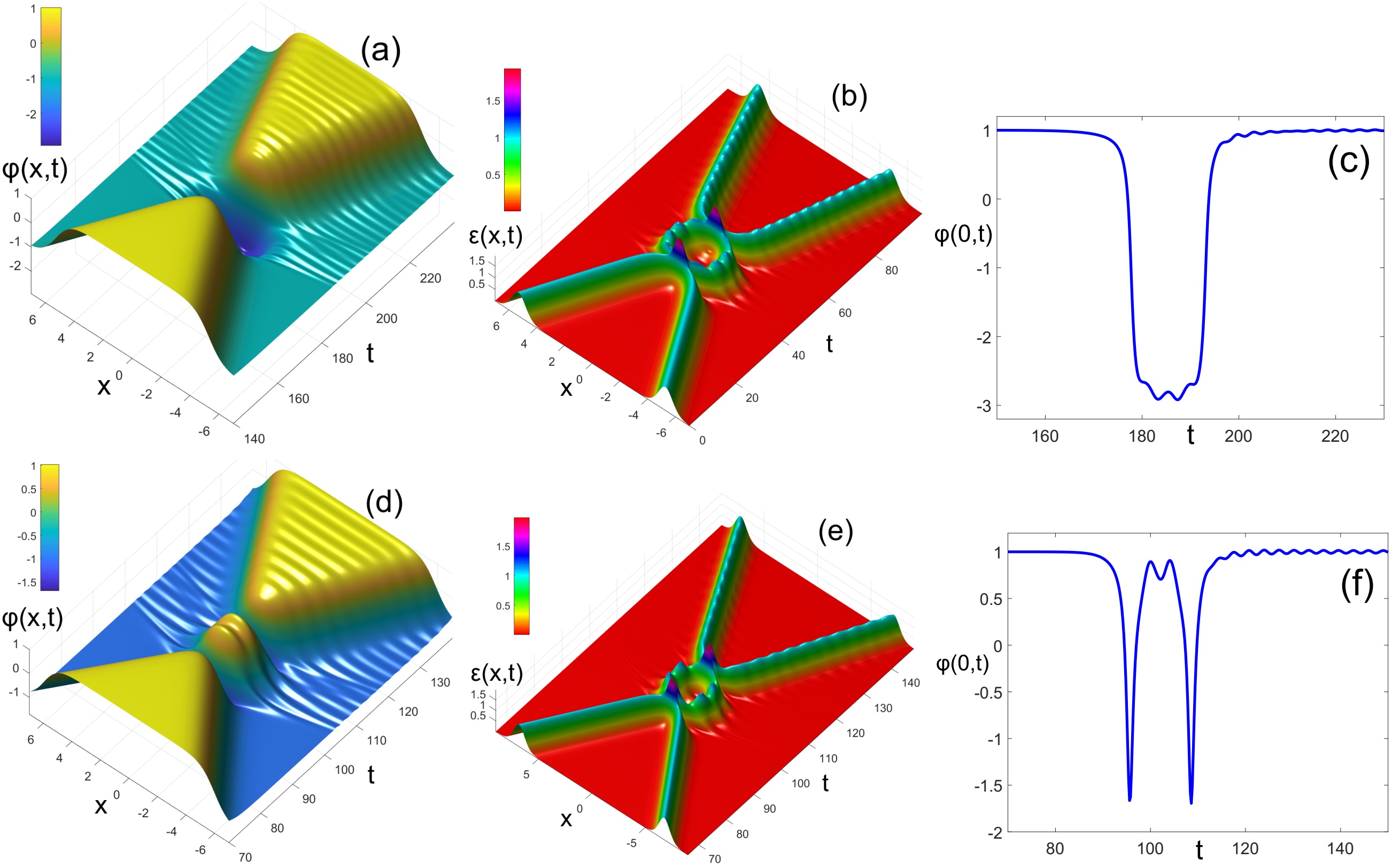}
   \caption{ The second (first)  row shows   a two-bounce resonance in the context of the $\varphi^4$ (periodic  $\varphi^4$) system with the initial speed  $v=0.2$ ($v=0.105$). Figures $a$ and $d$  show the  field representations, Figs.~$b$ and $e$  show  the  energy density representation, and  Figs.~$c$ and $f$  show  the variation of the fields at the center of mass $\varphi(0,t)$ for these collisions.} \label{RW}
 \end{figure}

For many  systems, including  the  $\varphi^4$  and periodic  $\varphi^4$ system,
there have been numerous wide and narrow intervals of the initial speed  below $v_{c}$,
instead of forming a bion, kink and antikink finally escape after finite times of collisions.
For wide intervals, they usually collide for the first time, lose their kinetic energy and generate a  bion state that immediately turns into a pair of kink-antikink near the collision point.  They collide for the second time generating another bion state which leads to a pair of kink-antikink that  gets separated and travel back to their starting points. This phenomenon is known as the two-bounce
resonance \cite{EZ00,FR2,FR1}.
Thus, the  interesting part is finding  intervals of the initial speeds  which lead to two-bounce
resonances, such a special interval is called a two-bounce (scattering)  window. In  Fig.~\ref{RW}, for instance, a two-bounce resonance is  shown for systems $\varphi^4$ and periodic $\varphi^4$, respectively.

\begin{figure}[ht!]
   \centering
   \includegraphics[width=150mm]{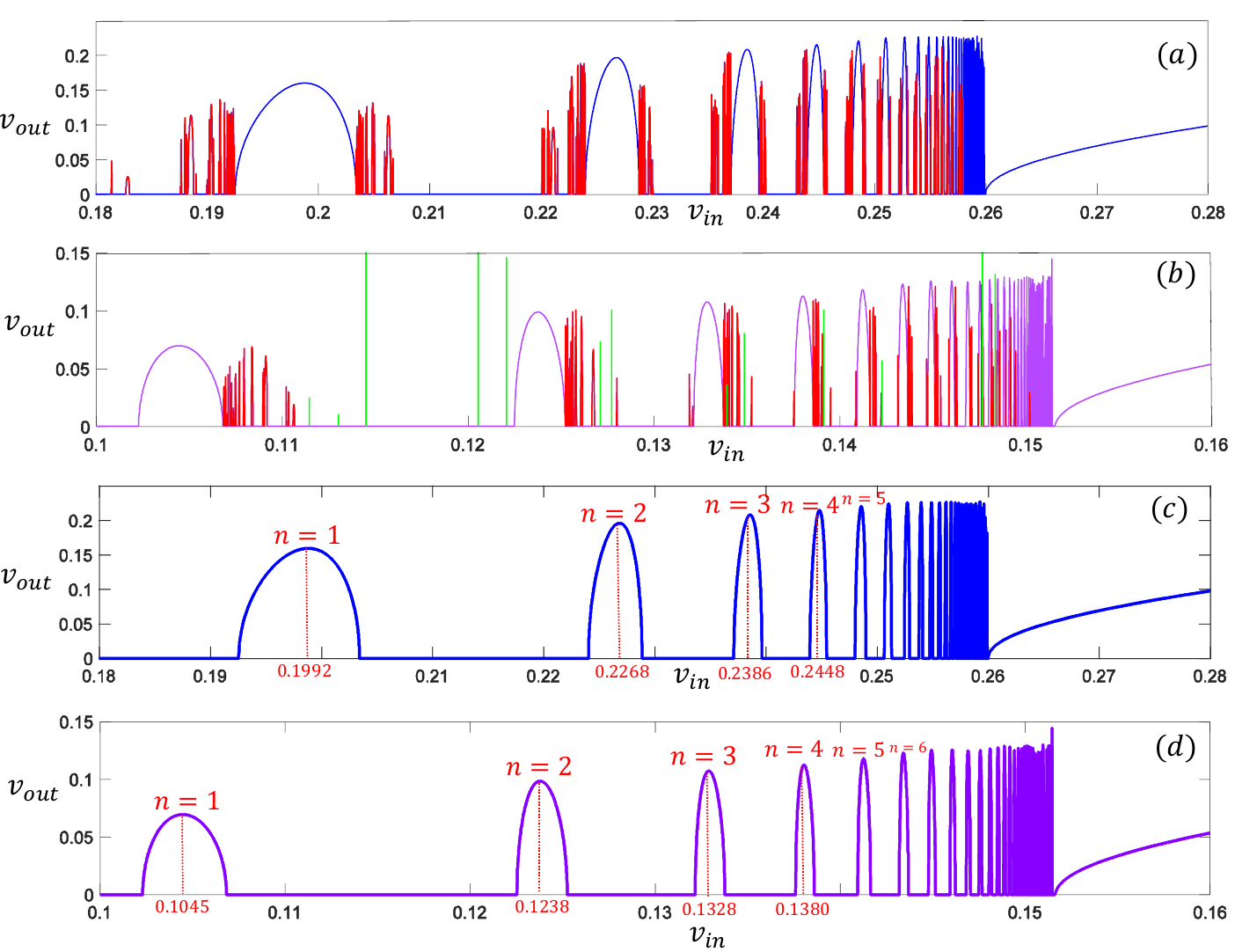}
   \caption{The output velocity versus  the input velocity.
   Figures  $a$ and $c$ ($b$ and $d$) belong to $\varphi^4$ (periodic $\varphi^4$) system.  Blue and purple peaks are the  two-bounce resonance windows and they can be labeled by incremental integers from left to right.     The other red sharp peaks are usually the three-bounce  resonance windows which  accumulated almost symmetrically (asymmetrically) around the two-bounce resonance windows of the $\varphi^4$ (periodic $\varphi^4$) system.
    The very sharp green peaks in Fig.~$b$ demonstrate  the situations for which the kink-antikink  collision eventually leads to a pair of bion states. } \label{twobounces}
 \end{figure}

\begin{figure}[ht!]
   \centering
   \includegraphics[width=150mm]{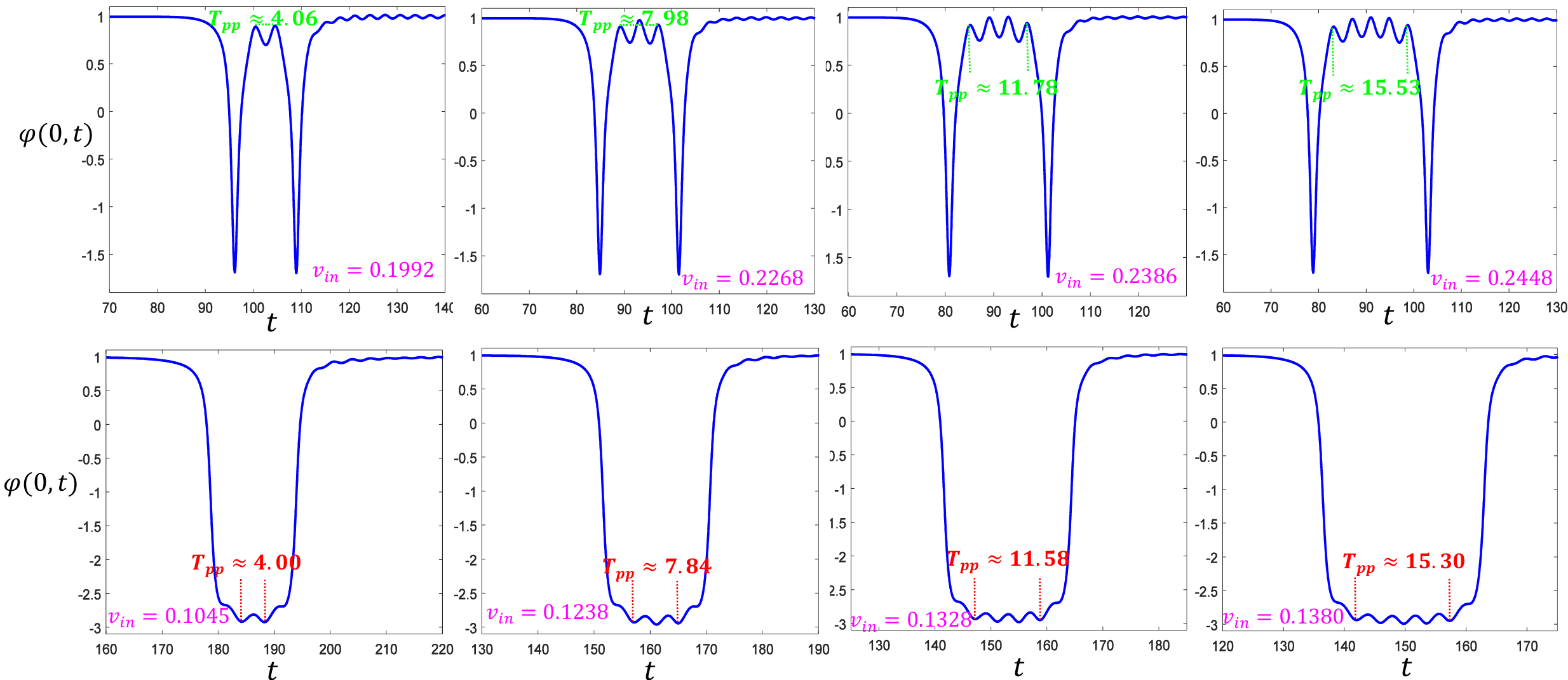}
   \caption{The variation of the field at the center of mass of a kink-antikink collision for different initial velocities belong to  first, second, third, and fourth two-bounce windows. First (second) row belongs to the  $\varphi^4$ (periodic  $\varphi^4$) system. The ($n+1$)th two-bounce    window differs from ($n$)th two-bounce    window by a longer time interval between  its first and second collision, and an additional cycle oscillation. The peak to peak time intervals are introduced in these  figures. } \label{cycle}
 \end{figure}

\begin{figure}[ht!]
   \centering
   \includegraphics[width=100mm]{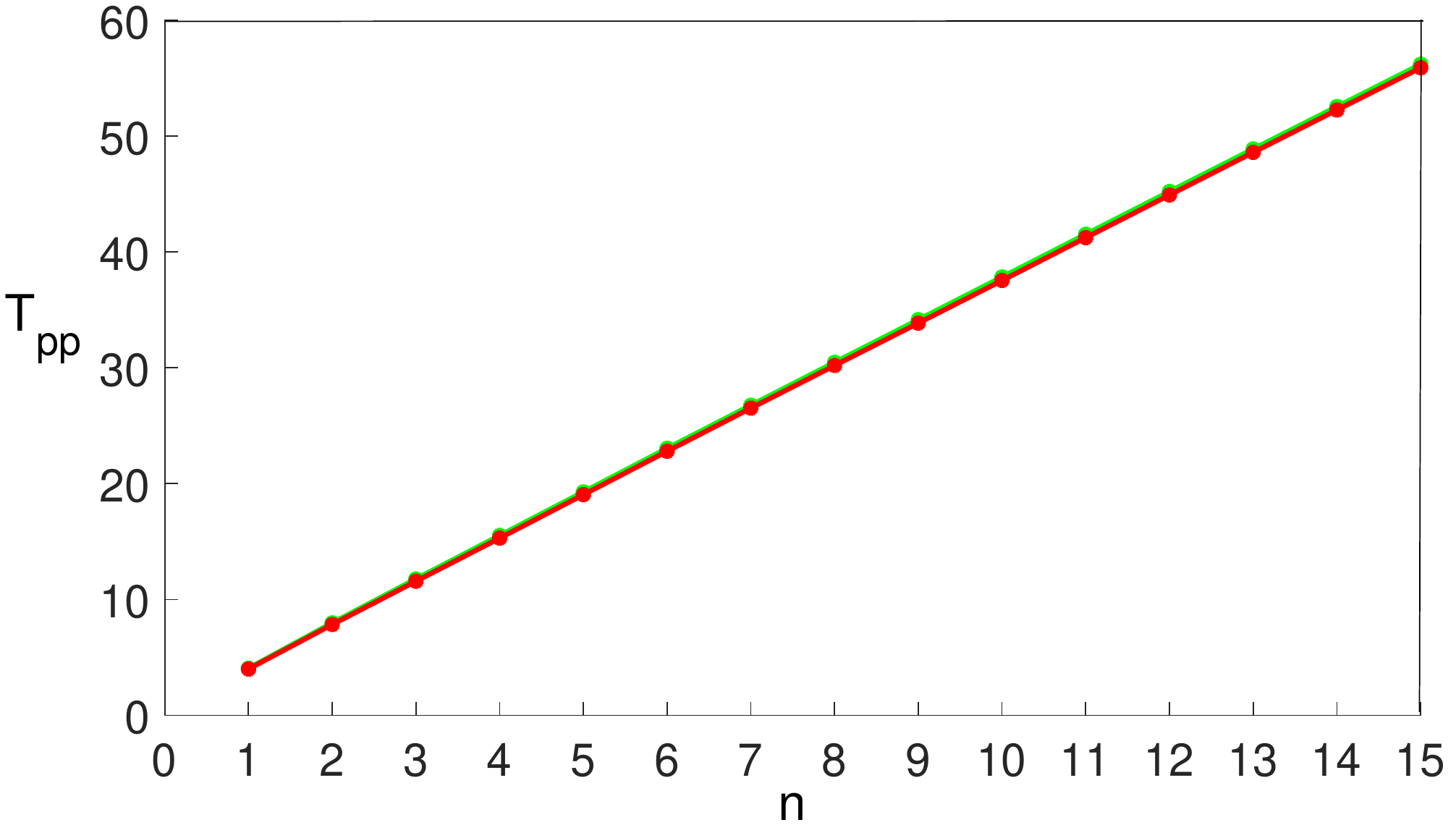}
   \caption{The peak to peak time intervals $T_{pp}$ (see Fig.~\ref{cycle}) versus the order of different two-bounce scattering windows (see label  $n$ in Fig.~\ref{twobounces}). The green (red) dots show the results of our numerical calculations for the  $\varphi^4$ (periodic  $\varphi^4$) system. Although the two systems have different potentials for regions  $\varphi<-1$, and $\varphi>1$ and different $v_{out}-v_{in}$ diagrams, interestingly they exhibit almost similar behavior in relation to the time elapsed between successive collisions.} \label{timecycles}
 \end{figure}

\begin{figure}[ht!]
   \centering
   \includegraphics[width=150mm]{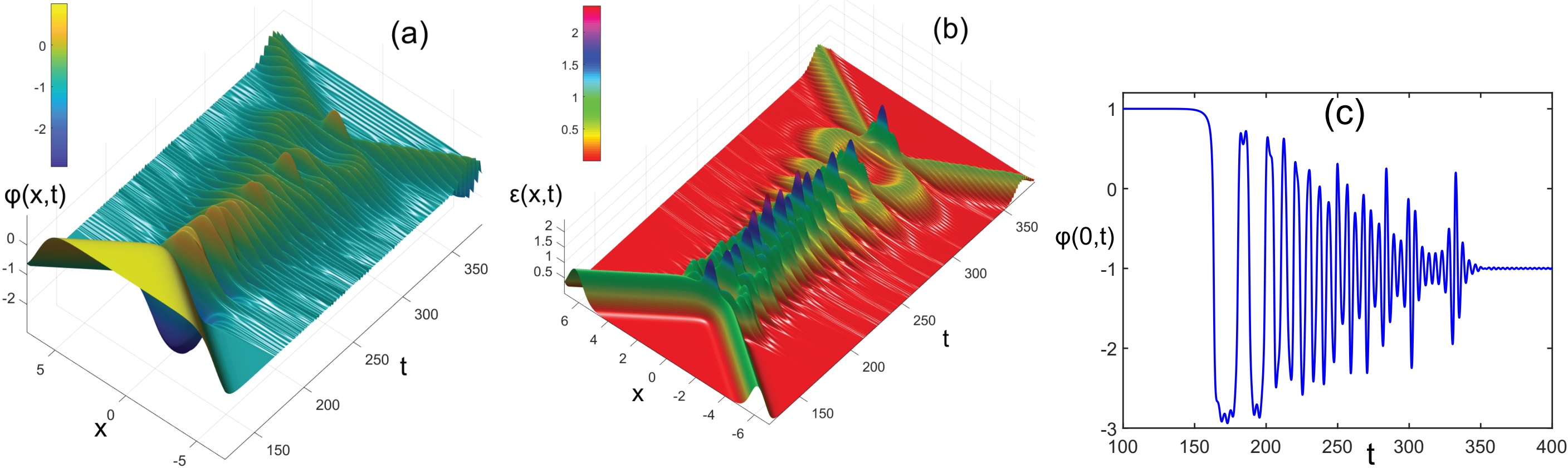}
   \caption{Generation of a pair of bion states in a kink-antikink collision of the periodic $\varphi^4$ system with the initial speed  $v=0.11454$. This is not the case for the $\varphi^4$ system.} \label{bion}
 \end{figure}

For better understanding, for many kink-antikink  collisions,  we prepared the output   speed as a function of  the  incoming  speed for the  $\varphi^4$ (periodic  $\varphi^4$) system  in the range from  $0.18$ to $0.28$ ($0.1$ to $0.16$) with the small step size  $0.00001$, numerically. The final result of the time-consuming computation is  Fig.~\ref{twobounces}-$a$ ($b$)  for the  $\varphi^4$ (periodic  $\varphi^4$)  system. We split the obtained  peaks into two groups of blue (purple) and red. The intervals where the blue (purple) peaks are located represent the   two-bounce windows of the  $\varphi^4$  (periodic  $\varphi^4$) system. Different blue (purple) peaks corresponding to different two-bounce scattering windows   are counted from left to right depending on their position on the $v_{in}$-axis. Red peaks are usually another type of windows known as three-bounce windows that will be discussed subsequently.
In general, it seems that there are many discrete two-bounce windows that the width of them decreases as the initial speeds increase.  Numerical calculation  shows that the  $(n+1)$th two-bounce window differ from $(n)$th two-bounce window by a longer  time interval, corresponding  to   an additional  cycle oscillation between their first and second collisions \cite{EZ00} (see Fig.~\ref{cycle} and Fig.~\ref{timecycles}). In this regard, a peak to peak time interval ($T_{pp}$) can be introduced for the small  cycle oscillations  to be used as a criterion for comparing the time elapsed between the first and second collisions in  different two-bounce scattering windows  (see Fig.~\ref{cycle}). Figure~\ref{timecycles} illustrates the relation between  the peak to peak time interval ($T_{pp}$) and the number of different two-bounce windows ($n$). For both $\varphi^4$ and periodic $\varphi^4$ systems, this relation is  linear with a very good approximation.  Moreover, Fig.~\ref{timecycles} shows that the green and red dots  coincide  very well together, indicating that the two systems behave similarly for these small oscillations.
Furthermore, it should be noted  that only for the periodic $\varphi^4$  system, there are seen  another type of  very narrow detached intervals for which  two bion states reappear after  kink-antikink collisions (see Fig.~\ref{bion}), that is, the intervals correspond to the green peaks in Fig.~\ref{twobounces}-$b$.

%showing a critical velocity of about ($v_{cr} = 0.1516$) $v_{c} = 0.2600$ for (periodic) $\varphi^4$ system.

\begin{figure}[ht!]
   \centering
   \includegraphics[width=150mm]{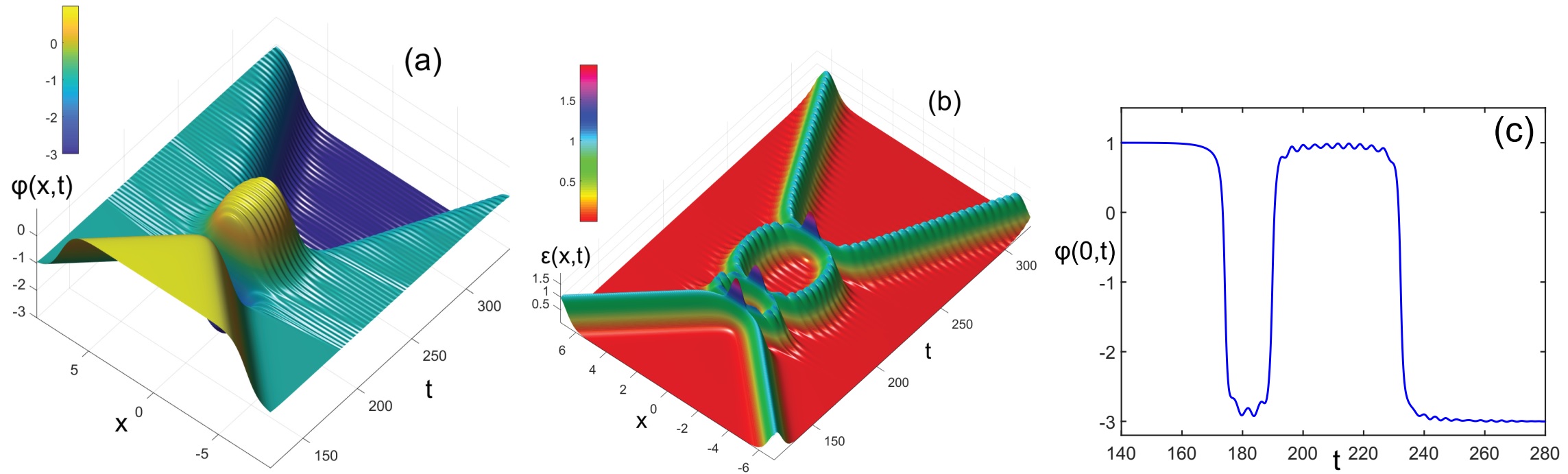}
   \caption{Three-bounce resonance in a kink-antikink collision of the periodic $\varphi^4$ system with the initial speed  $v=0.10722$.} \label{threeresonance}
 \end{figure}

\begin{figure}[ht!]
   \centering
   \includegraphics[width=120mm]{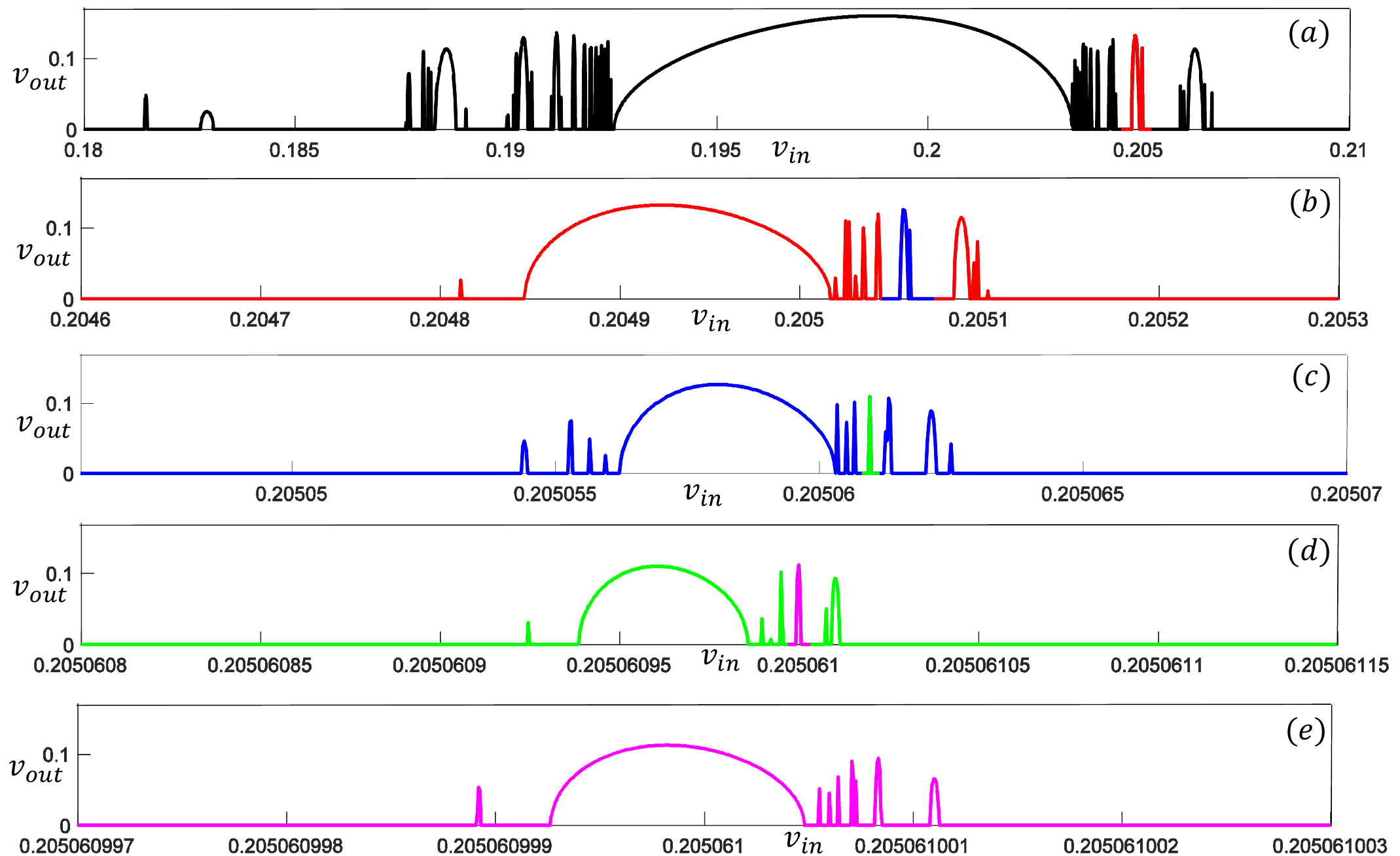}
   \caption{The quasi-fractal structure for the $\varphi^4$ system in $v_{out}$-$v_{in}$ diagram.} \label{phi4-zooms}
 \end{figure}

\begin{figure}[ht!]
   \centering
   \includegraphics[width=120mm]{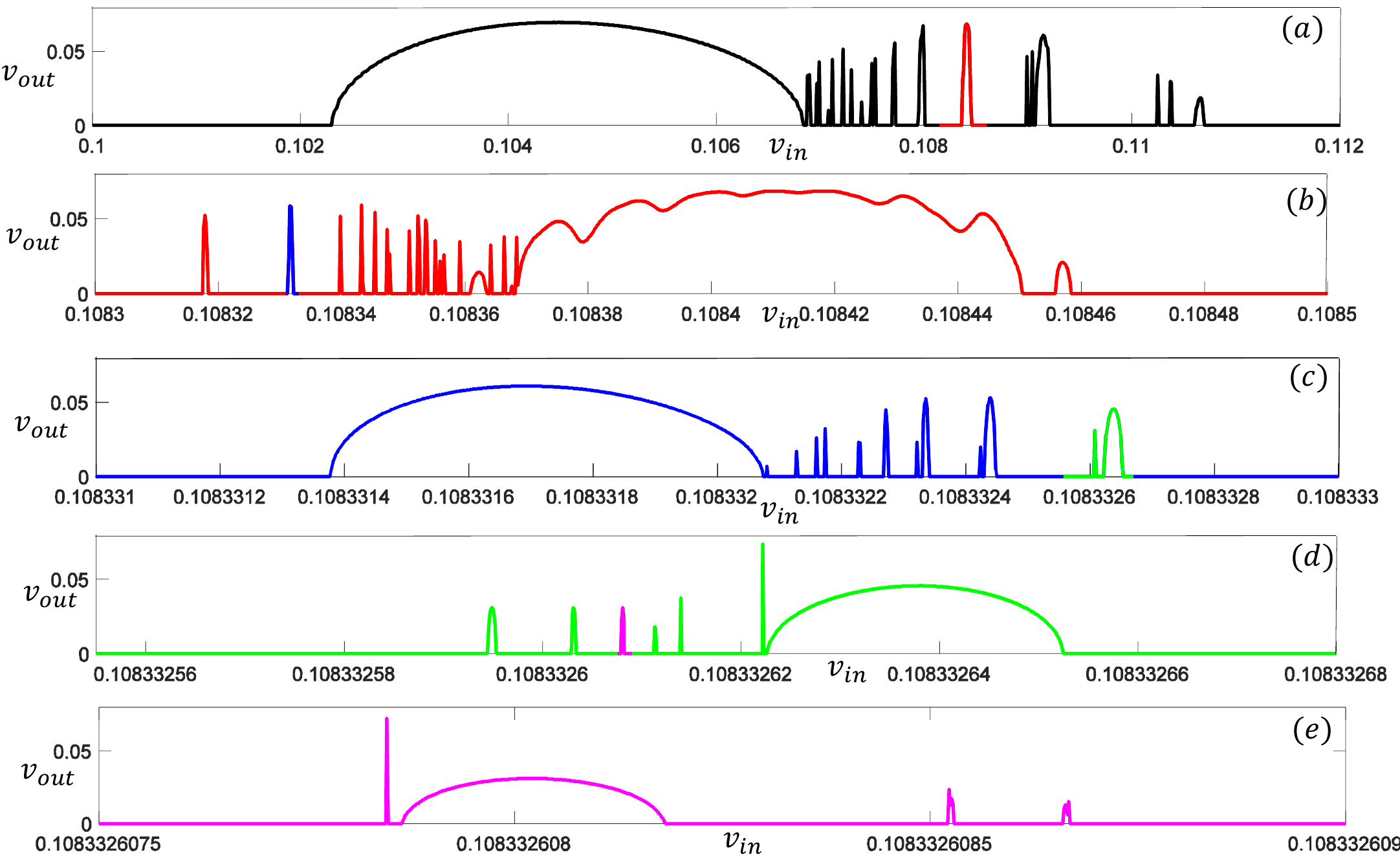}
   \caption{The quasi-fractal structure for the periodic $\varphi^4$ system in $v_{out}$-$v_{in}$ diagram.} \label{pphi4-zooms}
 \end{figure}

Around the wide blue (purple) peaks in Fig.~\ref{twobounces}, there are many  narrow red peaks, which indicate another type of scattering windows known as three-bounce scattering windows, in which kink and antikink collide three times and then  recede (see Fig.~\ref{threeresonance}). In general, there is a quasi-fractal structure for such peaks (corresponding to scattering windows), that is, for any narrow  $n$-bounce  window, corresponding to a sharp peak, there are some adjacent sharper  peaks which indicate a group of narrower  $(n+1)$-bounce windows. An $n$-bounce window is an interval of initial speed for which kink and antikink collide $n$ times before
bouncing back.
For example, zooming in on the first two-bounce window of the  $\varphi^4$ system in Fig.~\ref{twobounces}-$a$ in the range of $0.18$ to $0.21$, leads to Fig.~\ref{phi4-zooms}-$a$. Among the sharp  peaks around  the first two-bounce window,   we can consider a small  interval containing  a sharp peak, i.e.   the red part in   Fig.~\ref{phi4-zooms}-$a$ (for simplicity we refer  to it as  the red interval).
  If we divide the red narrow interval into $1000$ nodes,  and  perform  the numerical calculation  for all of them,  a more accurate $v_{out}-v_{in}$ diagram for  this interval   can be obtained (see Fig.~\ref{phi4-zooms}-$b$), which shows some new detail for the red small interval. Note that, since   Fig.~\ref{phi4-zooms}-$a$ is obtained by dividing the interval $0.18$ to $0.21$ into $300$ nodes, the red interval is then approximated  to only $7$ nodes,  which is not enough   to acquire   sharper  peaks around the red peak  in Fig.~\ref{phi4-zooms}-$a$ numerically.

Again, we can  choose  another small  interval containing  a sharp peak in Fig.~\ref{phi4-zooms}-$b$,  i.e.   the blue  region, and then a more accurate  diagram  with  $1000$ nodes  can be obtained  (see Fig.~\ref{phi4-zooms}-$c$). This routine was repeated for the small green and purple intervals in  next figures. Hence, the more we repeat the zooming in process, the sharper  peaks will be revealed.
What we can see here is the existence of  a quasi-fractal structure which can be considered as a general  rule for any  small interval containing a peak (in fact a n-bounce window), that is, we can see that there are other sharper peaks to the left and right each original peak. Similar results are noticed for all peaks in the $v_{out}-v_{in}$ diagram of  the periodic $\varphi^4$ system, as well (see Fig.~\ref{pphi4-zooms}).
  However, the more interesting rule is that if a special peak corresponds  to an $n$-bounce window, the  surrounding   left and  right peaks  are usually ($n+1$)-bounce windows. To give an example, the wide smooth purple peak in Fig.~\ref{pphi4-zooms}-$e$ is corresponding to a $6$-bounce scattering window and the three sharp peaks  around it are actually  $7$-bounce scattering windows.
The reported phenomena in the $\varphi^4$ and in the periodic $\varphi^4$ systems were studied before not only in the
$\varphi^4$  system \cite{EZ00,FR2,FR1}, but also in the modified sine-Gordon equation \cite{INTER1} and in the
double sine-Gordon model \cite{EZ1}.

% For the $\varphi^4$ system, it was shown that the edges of the resonance windows are
%not sharp regions and containing a  fractal hierarchial structure of n-bounce windows
%$n\geq 3$  [...]. Our numerical studies for the periodic $\varphi^4$ system yielded similar results for the edges of the resonance windows. However, there are some very sharp completely detached intervals from two-bounce intervals that the three-bounce resonance can be detected numerically Fig.~\ref{threeresonance}. Moreover, there are another type of  detached intervals for which  two bion state reappear after the kink-antikink collisions. For example, in the range of

%\begin{figure}[htp]
%
%  \centering
%
%  \label{fractal}
%
%  \begin{tabular}{cc}
%
%
%    \includegraphics[width=70mm]{theresonacewindows}&
%
%    \includegraphics[width=70mm]{resonancewindowsphi4comapring}\\
%
%
%  \end{tabular}
%\caption{The right (left) Fig is shown the output speed  as a function of the input speed for the  $\varphi^4$ (periodic  $\varphi^4$), and a critical speed  of about $v_{c} = 0.2598$ ($v_{c} = 0.15$) and several resonance windows. (Some details have been suppressed for clarity of exposition).}
%\end{figure}

It should be noted that in the figures obtained (\ref{phi4-zooms} and \ref{pphi4-zooms}), the more we zoom in, the narrower intervals are obtained  with  more decimal numbers. Hence,  the numerical results for such very narrow  intervals, due to the high sensitivity and inevitability of numerical errors, would  depend on  the type of space-time spacing.   Figures $d$ and $e$ were obtained for  $h = k = 0.02$ and may be changed  if we use other space-time spacing values. However, the original nature  of these systems does not change and is similar to what was seen in Figs.~\ref{phi4-zooms} and \ref{pphi4-zooms} by zooming in.

\subsection{Disturbed kink-antikink Collisions}

To study a disturbed kink-antikink collision with the initial velocities  $v$ and $-v$ and initial positions $a$ and $b$ (provided $|b-a|$ is large enough), for which at least one of the kink and antikink solutions get
excited, we first need to prepare the  initial conditions in the following form:
\begin{eqnarray} \label{dist}
&&\phi_{K\overline{K}}(x,t)=\tanh(+\gamma(x-vt-a))+\psi_{1}(\gamma(x-vt-a))\sin(\omega t-kx+\theta_{1})+
\nonumber\\&&
 \quad\quad\quad \quad\tanh(-\gamma(x+vt-b))+\psi_{2}(\gamma(x+vt-b))\sin(\omega t+kx+\theta_{2})-1,
\end{eqnarray}
where $\psi_{1}(x)$ and $\psi_{2}(x)$  are linearly dependent   small functions, $\theta_{1}$ and $\theta_{2}$ are two arbitrary initial phases  of  the kink and the antikink, respectively.
%Since $\psi_{1}$ and $\psi_{2}$ are small functions, according to Eq.~(\ref{varE}), the particle features   of the distinct kink and antikink at initial times (\ref{dist}) do not change significantly.
Although $\psi_{1}(x)$ and $\psi_{2}(x)$ are small and do not change the particle features   of the distinct kink and antikink at initial times  significantly,   it can be shown numerically that the  small trapped wave profiles  by kink and antikink, i.e. $\psi_{1}(\gamma(x-vt-a))\sin(\omega t-kx+\theta_{1})$ and $\psi_{2}(\gamma(x+vt-b))\sin(\omega t+kx+\theta_{2})$, have a crucial role  in the output of the collisions.  In fact the maximum amplitudes  of trapped wave-profiles  $A_{1}=\max{\psi_{1}}$ and $A_{2}=\max{\psi_{2}}$, and initial phases $\theta_{1}$ and $\theta_{2}$, are two important factors that can have a significant   impact on  the outcome of the collisions. Note that, the initial phases $\theta_{1}$ and $\theta_{2}$ are completely optional parameters  which can  be randomly    considered any amount in the initial conditions (\ref{dist}), however,  they  play an important role in the fate of a disturbed kink-antikink collision.

\begin{figure}[ht!]
   \centering
   \includegraphics[width=150mm]{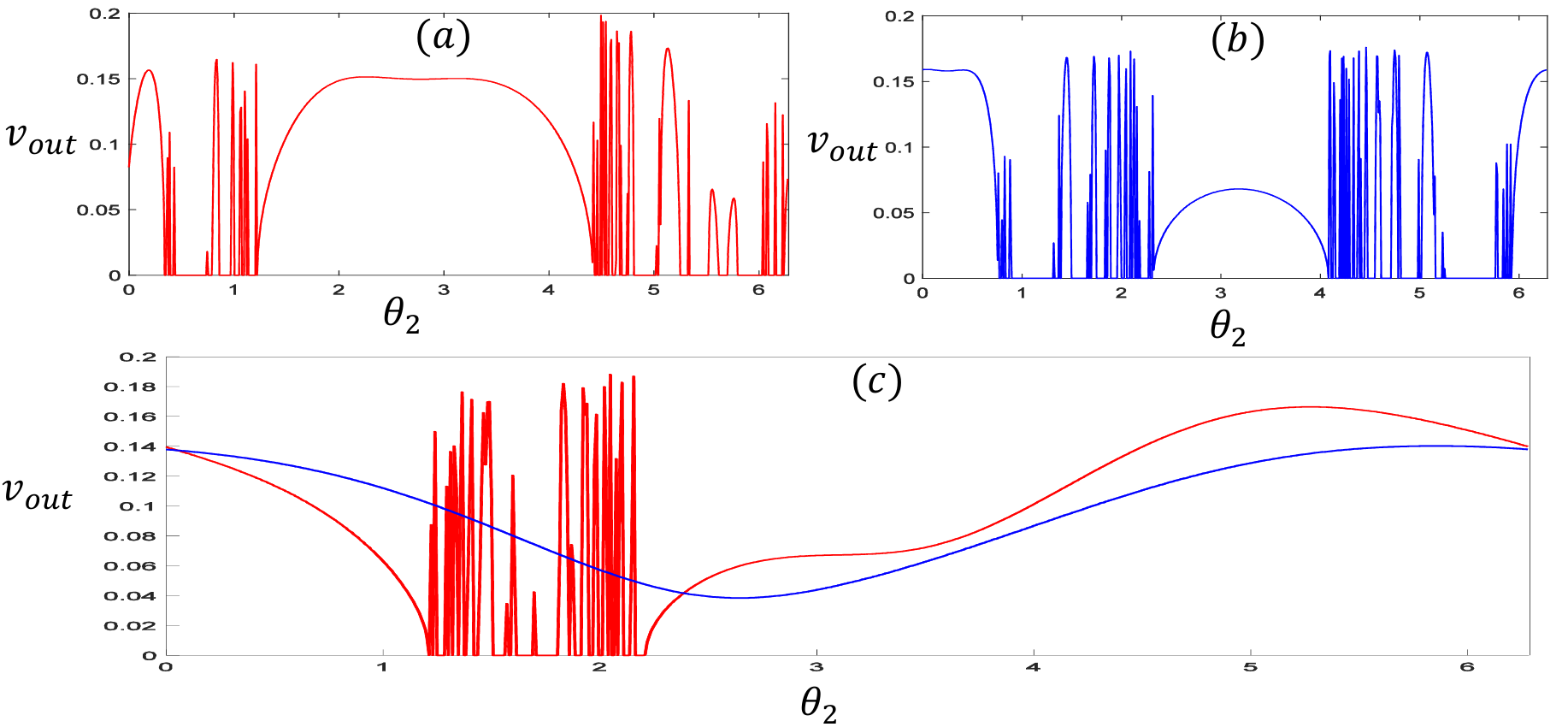}
   \caption{The output velocity versus $\theta_{2}$ for a disturbed kink-antikink  collision. Figs-$a$ and $b$ are obtained in  the context of the $\varphi^4$ system and Fig-$c$ is obtained in  the context of the periodic  $\varphi^4$ system.  For the red (blue) curves the  maximum  amplitude of the initial trapped wave profile is $A=0.1$ ($A=0.05$). Disturbed kink and antikink  initially  stand at $a=-20$ and $b=20$, the  initial speed is  $v=0.2$, and for disturbed kink $\theta_{1}=0$. To obtain these results,  we used $700$ nodes   for $\theta_{2}$ in the range from $0$ to $2\pi$.} \label{tetphi4}
 \end{figure}

\begin{figure}[ht!]
   \centering
   \includegraphics[width=150mm]{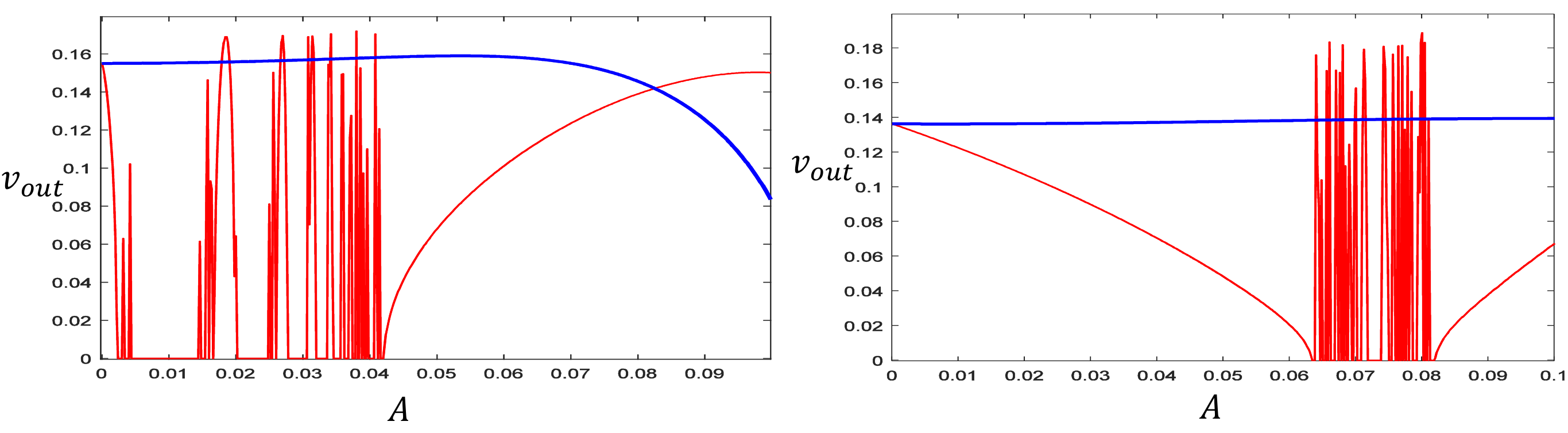}
   \caption{The left (right) figure presents the  output velocity versus $A=A_{1}=A_{2}$ for a disturbed kink-antikink  collision with $v_{in}=0.2$ in the context of the  $\varphi^4$ (periodic  $\varphi^4$) system. For the blue (red) curve the initial   phases are $\theta_{1}=\theta_{2}=0$ ($\theta_{1}=0$ and $\theta_{2}=\pi$). To obtain this figure we study all  collisions  in the range $0<A<0.1$ with   $500$ nodes.} \label{AAA}
 \end{figure}

For example, for  $\varphi^4$ and periodic $\varphi^4$ systems,  the red (blue) curves in Fig.~\ref{tetphi4} show how  the  output  speed of a kink  in a disturbed kink-antikink collision  with $b=-a=20$, $v=0.2$, $A_{1}=A_{2}=0.1$ ($=0.05$), and $\theta_{1}=0$, is  affected by   different optional choices of the initial  phase $\theta_{2}$.
Moreover, we obtain  the output velocity of the disturbed kink-antikink collisions versus the maximum  amplitudes of the initial wave profiles  $A_{1}=A_{2}=A$ in Fig.~\ref{AAA} for the $\varphi^4$ and periodic $\varphi^4$ systems, we set $\theta_{1}=\theta_{2}=0$ , $b=-a=20$, $v=0.2$.
Numerically, it was seen  that  high speed  collisions   (energetic collisions) reduce the influence of the initial trapped wave profiles  on the fate of collisions, i.e. we do not see significant different outcomes   in the outputs.

\begin{figure}[ht!]
   \centering
   \includegraphics[width=150mm]{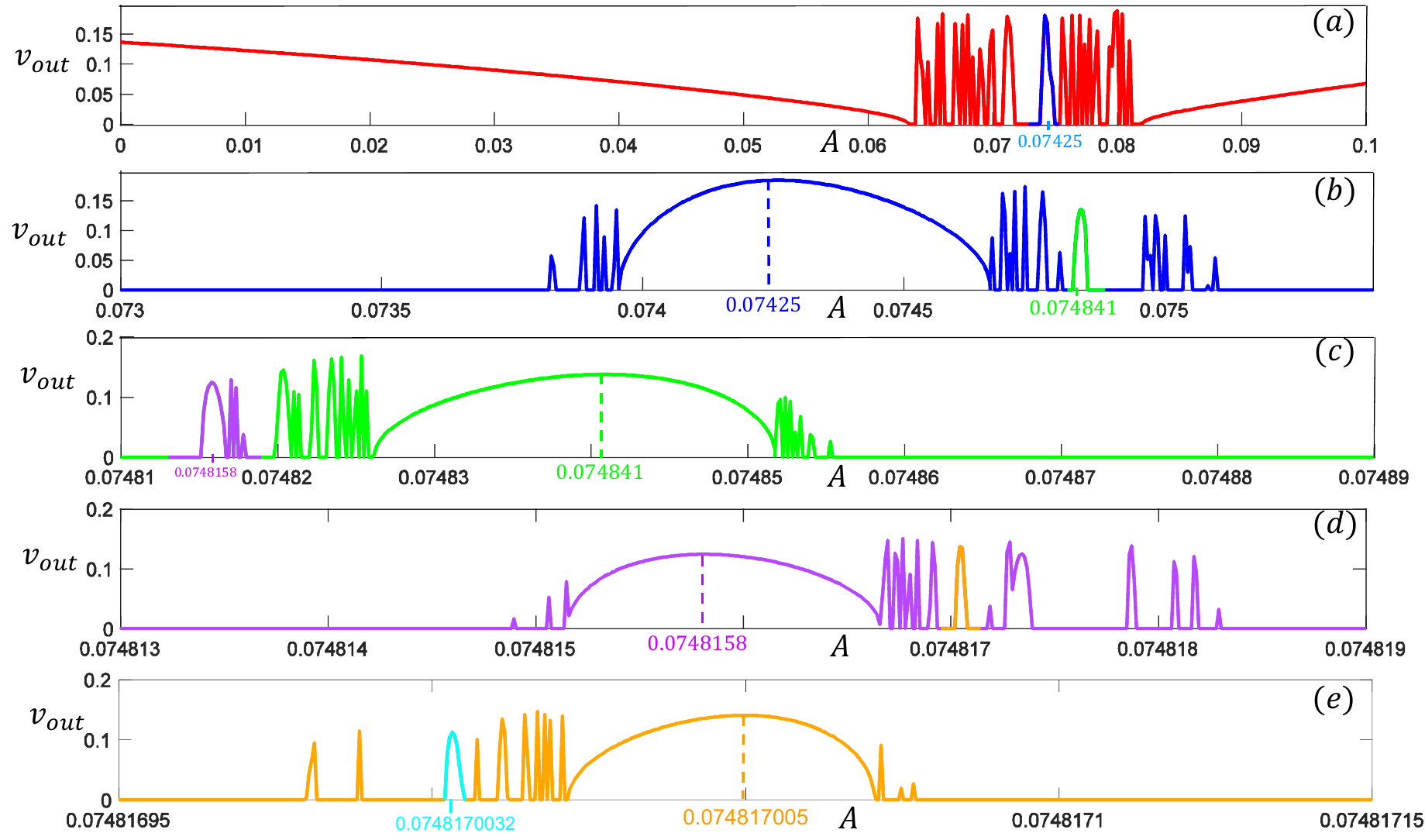}
   \caption{The quasi-fractal structure for the periodic $\varphi^4$ system in $v_{out}$-$A$ diagram. The blue peak in Fig-$a$, the green peak in Fig-$b$, the smooth part of the purple peak in  Fig-$c$, the orange  peak in Fig-$d$,  and the cyan peak in Fig-$e$ are related to a two-bounce, three-bounce, four-bounce, five-bounce, and a six bounce A-window, respectively. A similar explanation can be used for $v_{in}$-windows in Figs.~\ref{phi4-zooms} and \ref{pphi4-zooms}.} \label{blueredgreen}
 \end{figure}

In  Figs.~\ref{tetphi4} and \ref{AAA}, there are intervals for which  a  chaotic  behavior is seen. Similar to quasi-fractal structure of resonance windows which was discussed in the pervious subsection,  here   it was seen that the peaks in these special intervals  have a quasi-fractal structure as well. For instance,
we can consider  a small  interval containing  a sharp peak  in  the chaotic  region of the red curve in the  Fig.~\ref{AAA}-$b$, i.e.   the blue part in the  Fig.~\ref{blueredgreen}-$a$.
 According to Fig.~\ref{blueredgreen},  if we repeat the numerical simulation by $500$ nodes   just for the blue narrow  interval, a more accurate Fig.~(\ref{blueredgreen}-b) is obtained which shows  some new detail for the blue  interval. Again, we can choose another narrower   interval containing  a sharp peak in Fig.~\ref{blueredgreen}-$b$,  i.e.   the green part, and using a more accurate  simulation with  more nodes   to lead  to Fig.~\ref{blueredgreen}-$c$, and so on for the pink and orange small intervals in the next figures.

\begin{figure}[ht!]
   \centering
   \includegraphics[width=150mm]{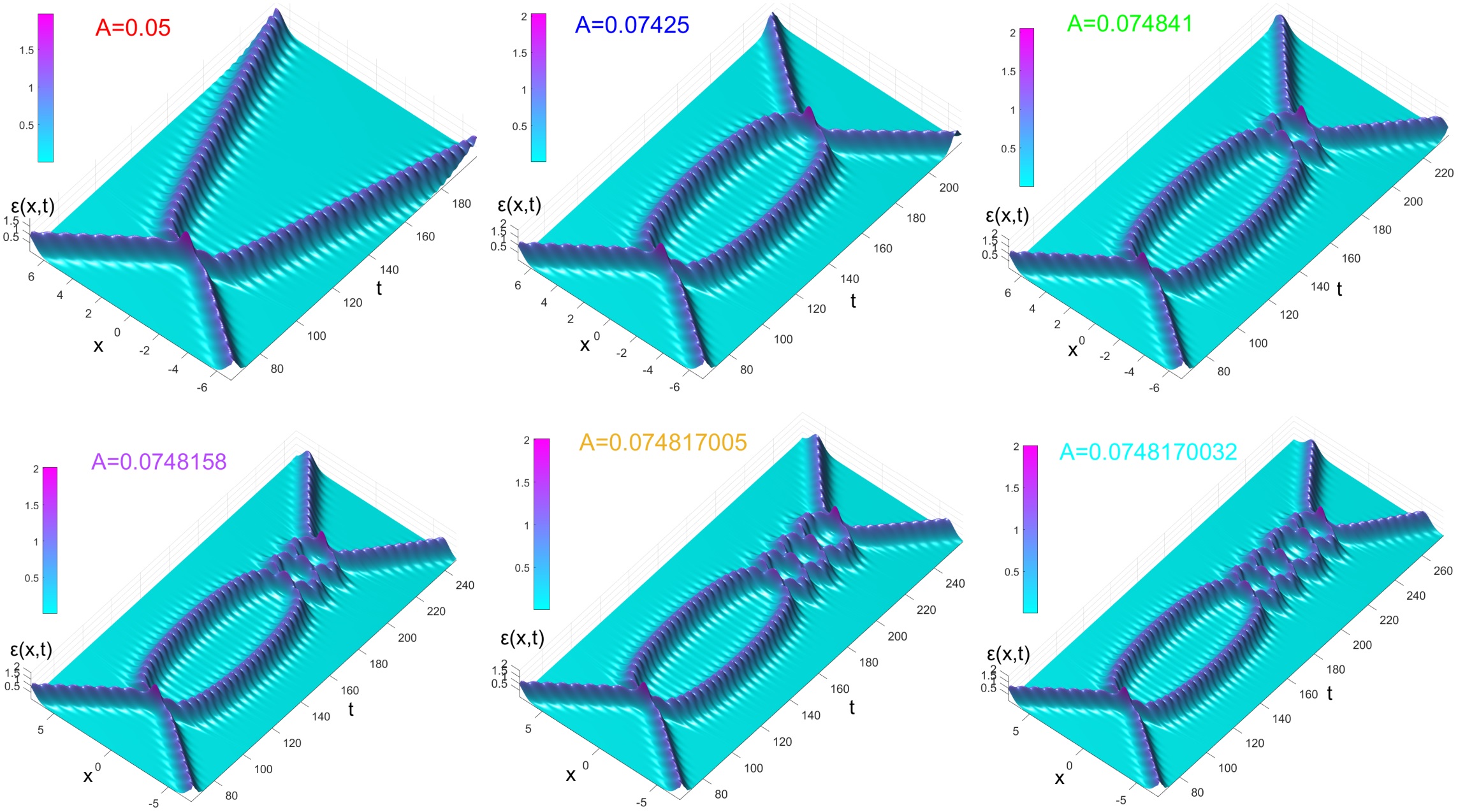}
   \caption{ The energy density representation of six disturbed kink-antikink collisions with different values of $A$ in the context of the periodic $\varphi^4$ system. We set  $v=0.2$, $\theta_{1}=0$, and  $\theta_{2}=\pi$. For different peaks ($A$-windows) in Fig.~\ref{blueredgreen},  different $n$-bounce ($n=2,3,4,5,6$) collisions would  occur. } \label{onetosixbounce}
 \end{figure}

\begin{figure}[ht!]
   \centering
   \includegraphics[width=150mm]{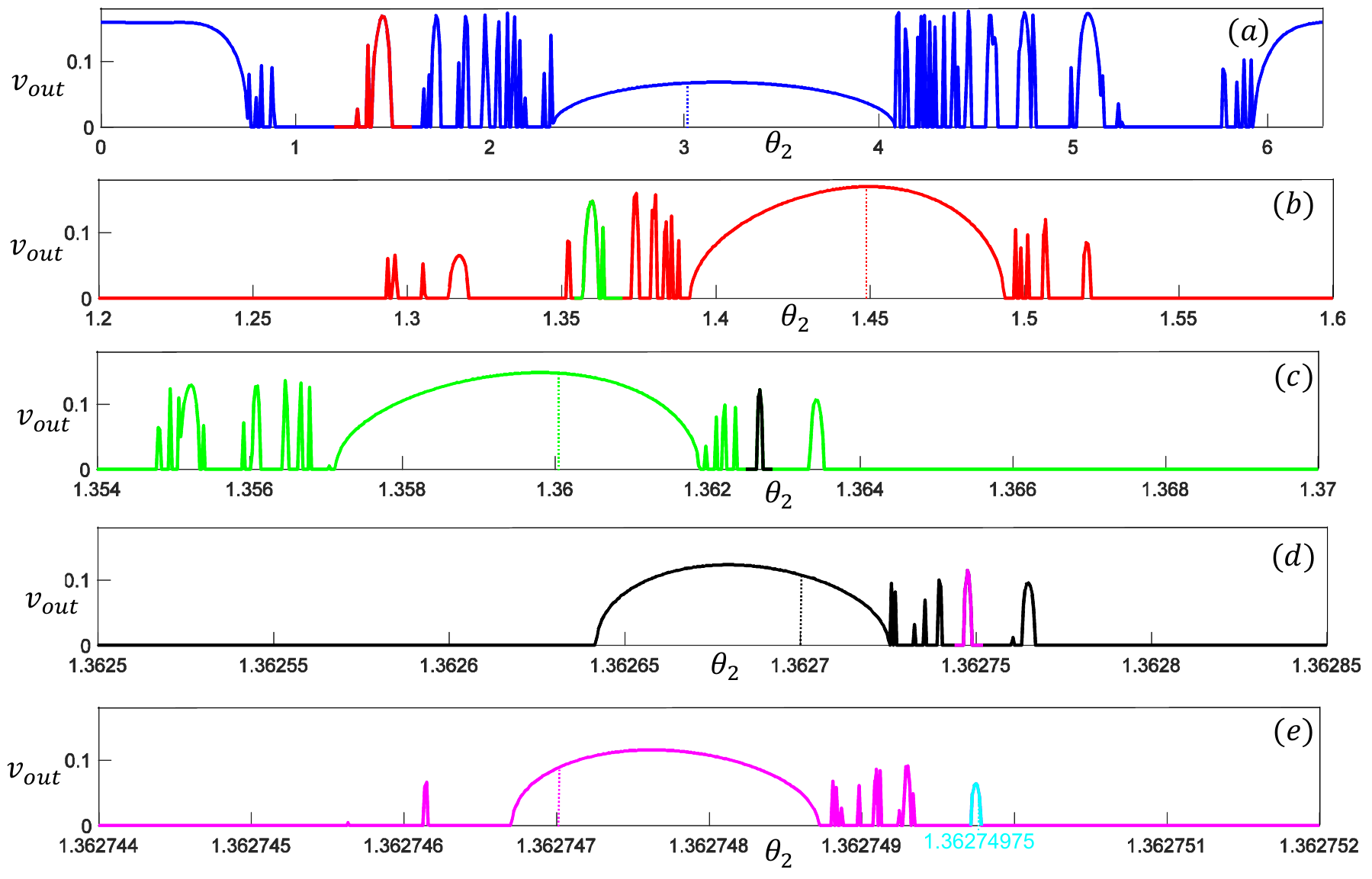}
   \caption{The quasi-fractal structure for the  $\varphi^4$ system in $v_{out}$-$\theta_{2}$ diagram. The blue peak in Fig-$a$, the green peak in Fig-$b$, the  black peak in  Fig-$c$, the purple  peak in Fig-$d$,  and the cyan peak in Fig-$e$ are related to a two-bounce, three-bounce, four-bounce, five-bounce and a six-bounce $\theta_{2}$-window, respectively. } \label{phase}
 \end{figure}

\begin{figure}[ht!]
   \centering
   \includegraphics[width=150mm]{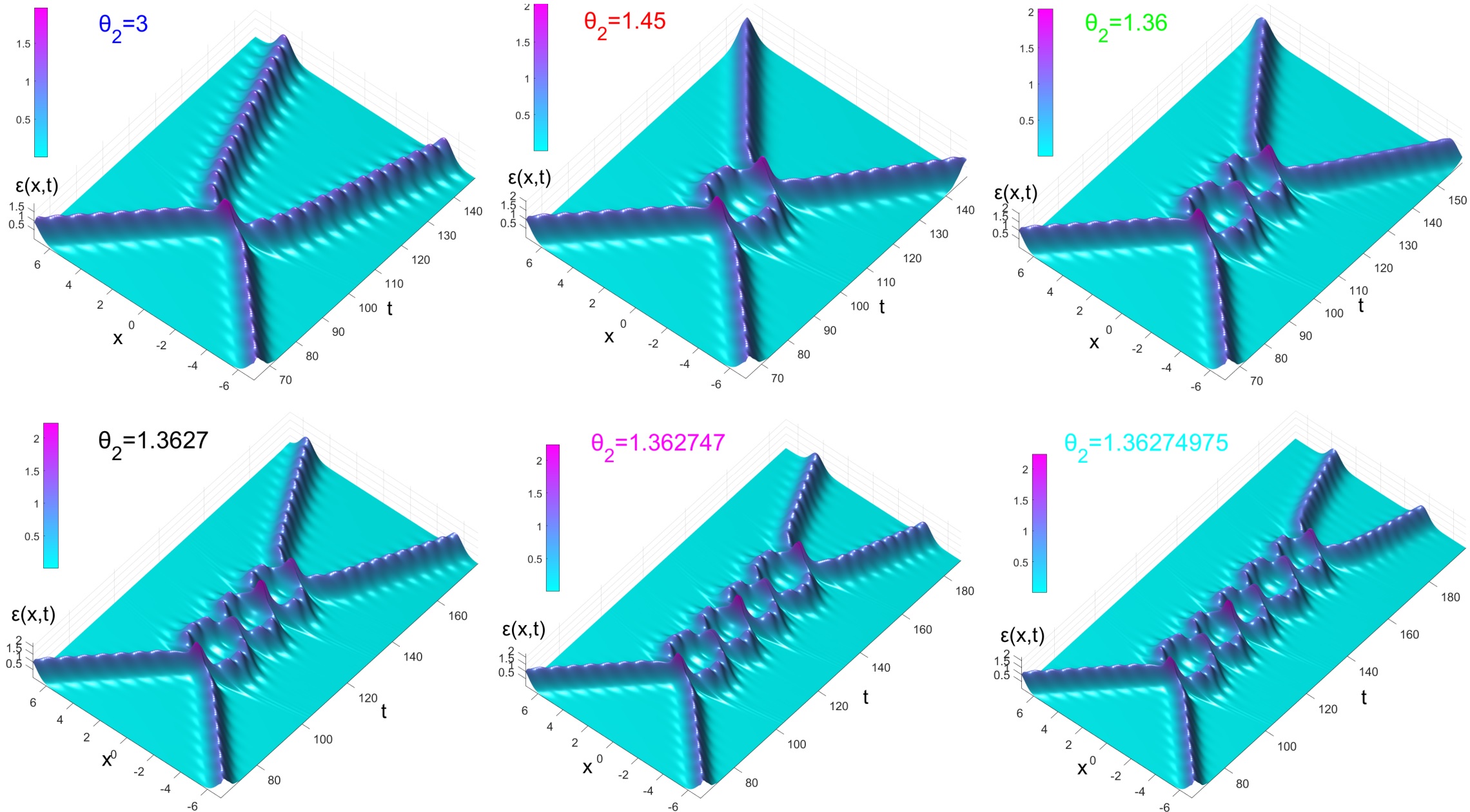}
   \caption{The energy density representation of six disturbed kink-antikink collisions with different values of $\theta_{2}$ in the context of the $\varphi^4$ system. We set  $v=0.2$, $\theta_{1}=0$, and $A=0.05$. For different peaks ($\theta$-windows) in Fig.~\ref{phase},  different $n$-bounce ($n=2,3,4,5,6$) collisions would  occur. } \label{phasewinows}
 \end{figure}

 What we can observe  here is the existence of  a quasi-fractal structure which can be considered to be a general  rule for any  small interval containing a peak  in any chaotic area. That is, as a rule, other sharper peaks surround each original peak from left and right. In the pervious subsection, to show  that there is a quasi-fractal structure, we studied  the outgoing velocity versus the incoming velocity to obtain the intervals with sharp peaks. Moreover, it was found that if any peak is an  $n$-bounce window,  the surrounding sharper peaks are usually  ($n+1$)-bounce windows.  A similar result is also obtained here for intervals with sharp peaks in the $v_{out}$-$A$ diagrams. In fact, for any sharp peak in the $v_{out}$-$A$ diagram,  we can define another type of  windows  with respect to parameter $A$ which can be called  an $A$-window. Studying the peaks in Fig~\ref{blueredgreen}~$a$-$e$, shows that
 if a special peak  corresponds  to  an $n$-bounce $A$-window,
 a substantial number of peaks surrounding it, are $(n+1)$-bounce $A$-windows
 (see Fig.~\ref{onetosixbounce}). Therefore, there is a similar  quasi-fractal structure in  chaotic regions of $v_{out}$-$A$ diagrams as well as the $v_{out}-v_{in}$ diagram.    These  results  can be generalized to the   $v_{out}$-$\theta$ diagrams as well.  To clarify, according to Fig.~\ref{tetphi4}-$b$,  we can select a special peak among the others  in Fig.~\ref{phase}-$a$ with the red color. Similar to the same process described in detail earlier, we show the results in Figs.~\ref{phase} and \ref{phasewinows} for $v_{out}-\theta_{2}$ diagrams.  For a more accurate analogy,  it may be better to call the  discussed windows in the $v_{out}-v_{in}$ diagrams, $v_{in}$-windows.

% Note that, since   Fig.~\ref{blueredgreen}-$a$ is obtained by dividing the interval $0\leq A\leq \pi$ into $500$ step sizes, then the blue interval is divided into just $5$ step sizes which are not enough   to obtain the surrounding  peaks  of the blue  peak  in Fig.~\ref{blueredgreen}-a numerically.

%
%What we can see here is the existence of  a quasi-fractal structure which can be considered to be a general  role for any  small interval containing a peak (in fact a n-bounce window) in any chaotic area. Therefore, as a role, we can see surrounding the left and the right of each peak, there are another sharper peaks. But the more interesting role is that if a special peak is a n-bounce window, its  major  surrounding the left and the right peaks are (n+1)-bounce windows. These  results  were  done just for this example in the  $\varphi^4$ (periodic  $\varphi^4$) system, but it can be generalized for any  chaotic  curve   in the  $\varphi^4$  and periodic $\varphi^4$ system.

%\begin{figure}[htp]
%  \centering
%  \begin{tabular}{cc}
%    \includegraphics[width=45mm]{teten1.pdf}
%    \includegraphics[width=45mm]{teten2.pdf}
%    \includegraphics[width=45mm]{teten3.pdf}\\
%
%    \includegraphics[width=45mm]{teten4.pdf}
%    \includegraphics[width=45mm]{teten5.pdf}
%    \includegraphics[width=45mm]{teten6.pdf}\\
%  \end{tabular}
%\caption{The energy density representation of a disturbed kink-antikink collision with $v=0.2$, $\theta_{1}=0$, and $A=0.05$ in the context of the $\varphi^4$ system. According to Fig.~\ref{phase}, we use six different $\theta_{2}$'s. Hence, belong to different peaks ($\theta$-windows) in Fig.~\ref{phase},  different $n$-bounce ($n=2,3,4,5,6$) collisions would  happen.  } \label{phasewinows}
%\end{figure}

\subsection{The Collision of  kink-antikink-kink}

\begin{figure}[ht!]
  \centering
    \includegraphics[width=145mm]{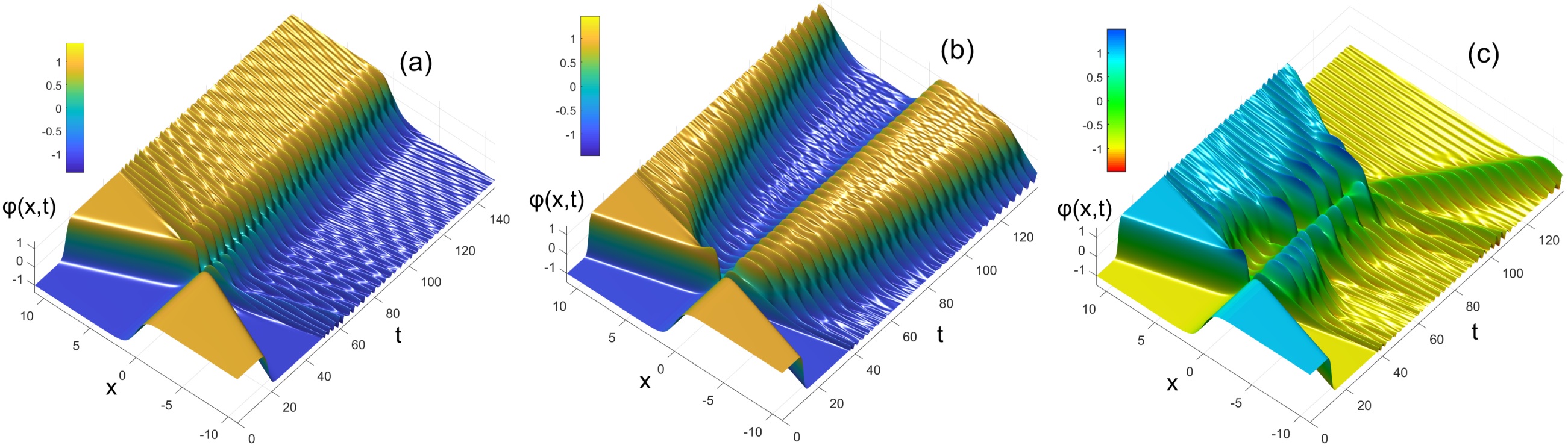}
\caption{The field representations of the $K\overline{K}K$ collisions in the context of the $\varphi^4$ system. For plots-$a$, $b$, and $c$, the initial speed is $v=0.5$, $v=0.765$, and $v=0.7591$, respectively.} \label{Threekinkphi4}
\end{figure}

\begin{figure}[ht!]
   \centering
   \includegraphics[width=150mm]{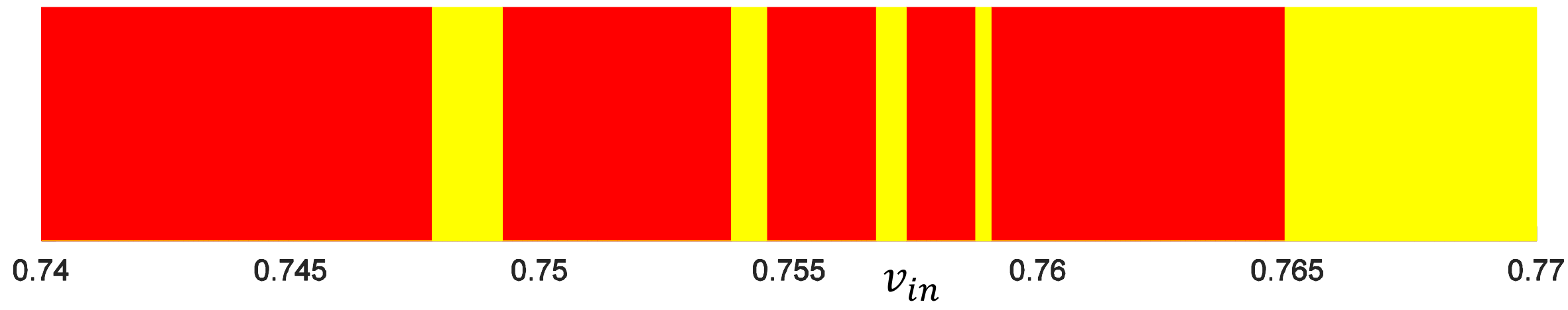}
   \caption{Different fates  versus  incoming speed  for  the $ K \overline{K} K $ collisions in the context of the $\varphi^4$ system.  The   red and yellow  bars  correspond to the intervals which leads to a single vibrating kink and   the reappearance of the triad   $K \overline{K} K $, respectively. In fact, four narrow  yellow bars indicate four scattering  windows.  To obtain this figure we studied  all  collisions  in the range $0.74<v_{in}<0.77$ with the step size of $0.0001$.} \label{colorphi4}
 \end{figure}

In this case we study the results of the collisions between  two kinks and one antikink  ($K\overline{K}K$). We set the initial conditions so that all the participants arrive at one point (origin) simultaneously.
The kinks, which are placed at $x_{3}=-x_{1}=20$, are moving towards the antikink placed at the origin ($x_{2}=0$) with the same speed.
The proper initial condition for this situation is:
\begin{equation}\label{jhp}
\varphi_{K\overline{K}K}=\tanh(+\gamma(x-vt-x_{1}))+
\tanh(-(x-x_{2}))+\tanh(\gamma(x+vt-x_{3})).
\end{equation}
In the context of the $\varphi^4$ system, the  critical  speed is about  $v_{cr}=0.7650$ for which if  $v<v_{cr}$, eventually a single at rest vibrating  antikink remains (Fig.~\ref{Threekinkphi4}-$a$), and if $v\geq v_{cr}$, the orientation  $ K \overline{K} K $ will reappear after the collision (Fig.~\ref{Threekinkphi4}-$b$).
 Only for the four  narrow  intervals of incoming speeds (yellow bars in Fig~\ref{colorphi4}-$b$),   close to the critical  speed, $ K \overline{K} K $ reappear after the collisions.  In fact, we can extend  the  concept of scattering   windows  for the $K\overline{K}K$ collisions, that is, there are some special intervals  of initial incoming velocities  slower than  the critical  speed for which $ K \overline{K} K $ can scatter each other, but they are always   two-bounce   scattering windows. At the edge of the scattering windows, another interesting phenomenon was observed which is  the appearance of   a vibrating antikink plus a bion state that  both leave the collision area in the opposite directions (Fig.~\ref{Threekinkphi4}-$c$).

%Studying the $K\overline{K}K$ collisions for the $\varphi^4$ system  is not rich that of the periodic $\varphi^4$ system.

\begin{figure}[ht!]
  \centering
    \includegraphics[width=145mm]{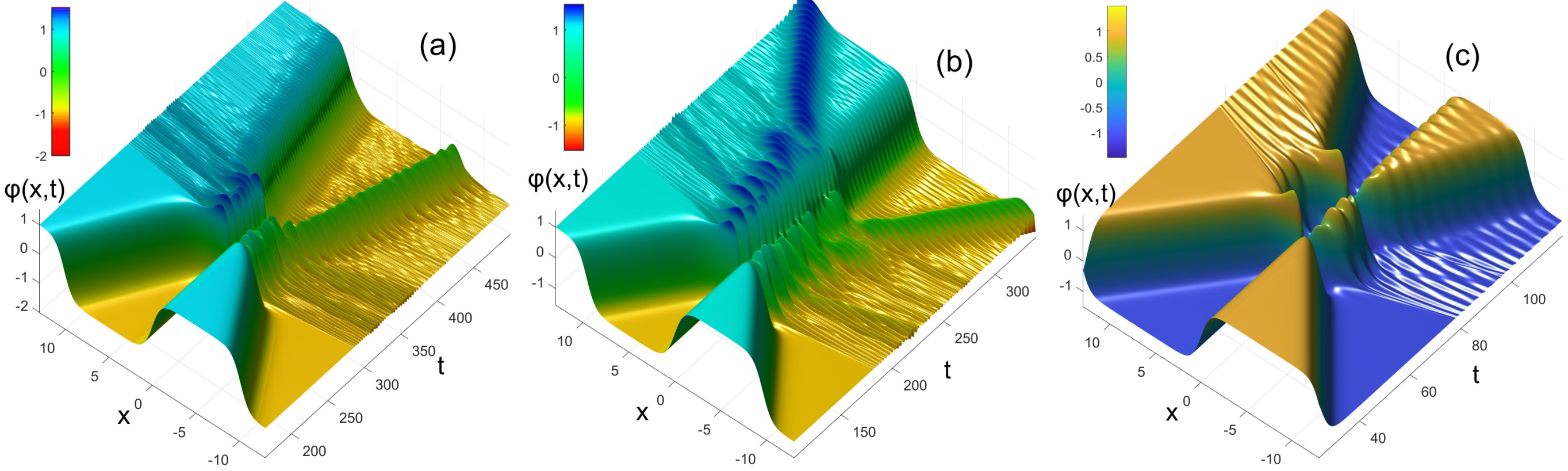}
\caption{The field representations of  $K\overline{K}K$ collisions in the context of the periodic $\varphi^4$ system. For plots-$a$, $b$, and $c$, the initial speed is $v=0.06$, $v=0.1$, and $v=0.28$, respectively.}  \label{Threekink}
\end{figure}

\begin{figure}[ht!]
   \centering
   \includegraphics[width=150mm]{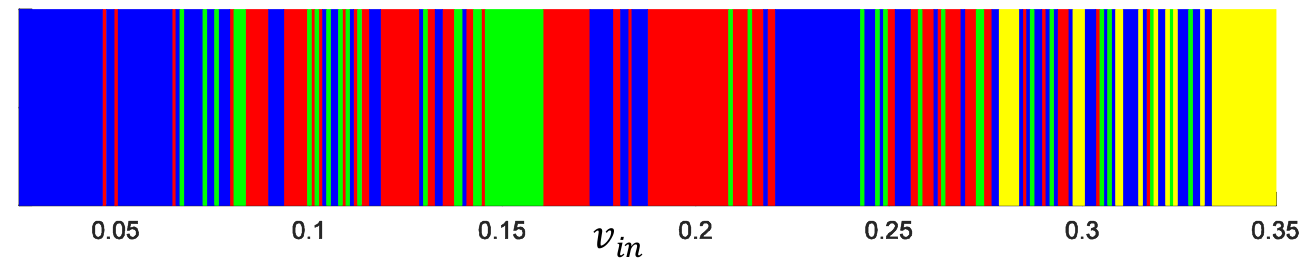}
   \caption{Variety  of  different fates  versus  incoming speed of the  $ K \overline{K} K $ collisions in the context of the periodic $\varphi^4$ system.  The blue, red, green, and yellow  bars are showing  different  intervals of incoming velocities which in turn lead to a bion state plus a vibrating kink, a single vibrating kink, two bion states plus a vibrating kink, and the reappearance of the triad   $K \overline{K} K $. To obtain this figure we studied  all  collisions  in the range $0.025<v_{in}<0.350$ with the step size of $0.001$.} \label{color}
 \end{figure}

For periodic $\varphi^4$ system, studying the $K\overline{K}K$ collisions leads to  different scenarios. Again, if the initial speed is higher than a  critical  speed of about $v_{cr}=0.3335$, the  orientation  $ K \overline{K} K $ will reappear after the collisions (see Fig.~\ref{Threekink}-$c$). Although, for the initial speeds smaller than $v_{cr}$, there are different intervals for the incoming speeds  with different outcomes.  More precisely,  for $K\overline{K}K$ collisions with $v<0.3335$, there are four different scenarios  at the end (see Fig.~\ref{color}). First,  a  vibrating antikink  and  a bion state remain and leave the collision area in the opposite directions (see Fig.~\ref{Threekink}-$a$ and the blue bars in Fig.~\ref{color}). Second, similar to Fig.~\ref{Threekinkphi4}-$a$, only a  standing vibrating   antikink remains (see the red bars in Fig.~\ref{color}).  Third, a  vibrating   antikink and two moving bion states remain (see  Fig.~\ref{Threekink}-$b$ and the green bars in Fig.~\ref{color}). Fourth, the triad $K\overline{K}K$  reappear after collisions (see  Fig.~\ref{Threekink}-$c $ and the yellow bars in Fig.~\ref{color}).  The forth case particularly characterizes the (two-bounce) scattering windows for the $K\overline{K}K$ collision in the context of the periodic $\varphi^4$ system.

%
%\begin{figure}[ht!]
%   \centering
%   \includegraphics[width=150mm]{Threekinkpphi4.jpg}
%   \caption{The field representations of  $K\overline{K}K$ collisions in the context of the periodic $\varphi^4$ system. For Figs-$a$, $b$, and $c$, the initial speed is $v=0.06$, $v=0.1$, and $v=0.28$, respectively.} \label{Threekinkpphi4}
% \end{figure}

Accordingly,  the study of three soliton-like  collisions  clearly shows us that despite the similarity of the   potential  in the range  from  $-1$ and $1$  for both systems (and then the similarity of the kink and antikink), the output of the collisions depends strongly on the potential in other ranges. In sum, the periodic $\varphi^4$ system provides a richer structure and more detail to study.

\section{Summary and Conclusion}

Based on the potential of the well-known relativistic kink-bearing $\varphi^4$ system,  we introduced a new  system that we call the periodic $\varphi^4$ system. The potential function in both systems is the same in the range $-1\leq\varphi\leq1$ which is corresponding to a kink (antikink) solution.
But for the periodic $\varphi^4$ system, we   use the same potential periodically in the range $-1\leq\varphi\leq1$ for other ranges from  $\pm1$  to $\pm$infinity. As long as we are dealing with a single kink (antikink),  everything is the same, and there is no difference between the two systems in terms of physical properties, but when it comes to  interaction with an antikink (a kink), due to   the potential difference in the other ranges, behaviors are practically different.
This paper attempts to provide a comparative study of the properties of two systems in interactions.

We have implemented a numerical program with proper accuracy in MATLAB, based on a fourth-order Runge-Kutta scheme to simulate the collisions (interactions).
We studied   the kink-antikink collisions  to  obtain  the scattering windows and other properties of  both systems.  For the  $\varphi^4$ (periodic  $\varphi^4$) system, the critical  speed is about $v_{cr}=0.2600$ ($v_{cr}=0.1516$). Usually for the speeds less than the critical  speed,  kink and antikink stick together and produce a bion state. However, in this range of  initial speeds, there are many wide and narrow intervals that the pair of  kink-antikink can scatter from one another. Such intervals on the axis of the incoming speed ($v_{in}$) are called the scattering windows (in this paper we call them $v_{in}$-windows).  The wide intervals in this range are two-bounce $v_{in}$-windows, and the surrounding  narrow intervals are $3$-bounce $v_{in}$-windows. For the two-bounce  $v_{in}$-windows, it was seen numerically that   the time interval between the first and second collisions   increases in a linear fashion versus the number of  successive two-bounce $v_{in}$-windows approximately in the same way  for both systems.
 Around the three-bounce $v_{in}$-windows, there  are some narrower intervals which are $4$-bounce $v_{in}$-windows, and so on, that is, there exists quasi-fractal structures for both $\varphi^4$ and periodic $\varphi^4$ systems in $v_{out}-v_{in}$ diagram where  $v_{in}<v_{cr}$.
  The notable difference in  $v_{out}-v_{in}$ diagrams is that  the peaks around  the two-bounce  $v_{in}-$windows in the $\varphi^4$ system are symmetrically distributed to the left and right, but for the periodic $\varphi^4$ system, they mainly appear to the right of them.
Furthermore, another interesting phenomenon in studying the kink-antikink collisions  of the periodic $\varphi^4$ system (that is not the case of the $\varphi^4$ system)  is the existence of very narrow intervals of the initial speeds for which a pair of bion states  appear as a result of a kink-antikink collision. This  phenomena were also  observed in \cite{EZ11,EZ16,EV1}, that called “\emph{a bound state of two oscillations}".

% If one studies  the  output speed of a kink-antikink as a function of the incoming speed,  a  threshold speed would be obtained for which if the incoming speed is being larger than that always a pair of vibrational kink-antikink reappear.

Since the kink and antikink in  $\varphi^4$ (periodic  $\varphi^4$) system have a non-trivial internal mode, they can get excited  and  have a constantly internal vibrational motion after each collision. This internal motion can be specified by two parameters: amplitude of the trapped wave profile  $A$ and the initial phase $\theta$. For the same initial speeds larger than the critical  speeds, the amplitude  of the imposed internal vibrations on the kink and antikink after the collisions,    in the context of the  periodic $\varphi^4$ system, are smaller than that of  the $\varphi^4$ system, thus,  the output  speeds are higher in value in the  periodic $\varphi^4$ system.   In a disturbed kink-antikink collision, for which at least one of the kink and antikink get excited, the study of  $v_{out}$-$A$ and $v_{out}$-$\theta$ diagrams at a constant incoming speed show that we can introduce other types of scattering windows on the axis of $A$ and $\theta$, which can be called $A$-windows, and $\theta$-windows. For such windows, we noticed a quasi-fractal structure similar to $v_{in}$-windows.

By looking at Figs.~\ref{phi4-zooms} ($a$-$e$), \ref{pphi4-zooms} ($a$-$e$), \ref{blueredgreen} ($b$-$e$), and \ref{phase} ($b$-$e$), it seems there  is a general  rule for the fractal structure in   $\varphi^4$ and periodic $\varphi^4$ systems. In fact, in the context of the $\varphi^4$ (periodic  $\varphi^4$) system,  for any wide peak, if the selected sharp peak (which is identified by a different color) is on the right of that, in the next step,  the surrounding sharper peaks are mainly on the right (left). The same statement can be reused by replacing the words left and right. For example, in Fig.~\ref{blueredgreen}-$c$ which was obtained for the  $\varphi^4$ (periodic  $\varphi^4$) system, we select a sharp purple peak on the left of the wide green peak, then in the next Fig.~\ref{blueredgreen}-$d$, the main surrounding sharper peaks are  on the right. As another example, in Fig.~\ref{phase}-$d$, the selected sharp purple peak on the right side  of the wide black  peak, leads to a wide purple peak with some surrounding sharper peaks mainly located on the right side of that.

Considering the collisions of three solitons ($K\overline{K}K$) in both systems and comparing them, show that the diversity  of  phenomena  in the periodic $\varphi^4$  is  richer  than the $\varphi^4$ system. For both systems, there are different critical speeds, that  $K\overline{K}K$ always scatter from each other and reappear after the collisions with $v>v_{cr}$.
In the context of the periodic $\varphi^4$ system,  the  collision  of the  $K\overline{K}K$ for $v<v_{cr}$ occur in four different  scenarios: First, the appearance of a bion state plus a vibrating kink. Second, a single vibrating kink remains after collision. Third, two bion states plus a vibrating kink appear after collision. Forth,  the reappearance of the triad   $K \overline{K} K $ which specifies the scattering  windows. However, in the context of the   $\varphi^4$ system for $v<v_{cr}$, we can also see narrow intervals close to $v_{cr}$ for which the $K\overline{K}K$ reappear after collisions. Otherwise, the    $K \overline{K} K $ collision always leads to  a single oscillating kink. Near the edge of the scattering windows in the $\varphi^4$ system, the appearance of a bion state plus a vibrating kink can  occur exceptionally.

%p134

 %The case four would  happen  for the incoming speeds larder than the critical speed, but for speeds slower than the  critical speed there are some intervals for which the case four happens, we can call $v_{in}$-windows for $K\overline{K}K$ collisions.

Although both $\varphi^4$ and periodic $\varphi^4$ systems have the same form of potential in the range $-1\leq \varphi\leq 1$ and  have  the same kink and antikink solutions,  their differences elsewhere  (i.e. $\varphi<-1$ and $1<\varphi$) will cause significant changes in the interactive features. Hence, we can call the form of  potential  elsewhere   “\textit{ interaction potential}''. This idea can be used to introduce  any other type of  $\varphi^4$ systems with different interaction potential forms.   For example, one can study a modified $\varphi^4$ system with a interaction  potential  in the following form:
\begin{equation}\label{fje}
V(\varphi)=\frac{1}{2}B(\varphi^2-1)^2, \quad \varphi<-1\quad \textrm{and} \quad 1<\varphi,
\end{equation}
where $B$ can be any arbitrary positive number. It is undeniable that the case $B=1$ is the same ordinary $\varphi^4$ system. Many interesting features, such as $v_{cr}$, can be obtained as a function of the parameter $B$. Furthermore, one can study the  well-known  SG system with different interaction potential forms as well.

%\subsection{The Collision Of Four Kinks}
%\subsubsection*{The Configuration (0,1,0,1,0)}

%\begin{figure}[h!]
%\centering
%\includegraphics[scale=1]{pic}
%\caption{The total energy density.
%	The kinetic energy density.
%	The gradient energy density 0.2 velocity.}
%\end{figure}

%\bibliographystyle{acm}

\end{document}